\documentclass[aps, prd, twocolumn, nofootinbib,showpacs]{revtex4}

\usepackage{bm}
\usepackage{graphics}
\usepackage{amsmath}
\usepackage{amssymb}
\usepackage{amscd}
\usepackage{afterpage}
\usepackage{epsfig}
\usepackage{psfrag}

\newcommand{\leftangle}{\langle\thickspace}
\newcommand{\rightangle}{\thickspace\rangle}
\newcommand{\lvertical}{| \thickspace}
\newcommand{\rvertical}{\thickspace |}
\newcommand{\half}{\frac{1}{2}}
\newcommand{\midle}{\thickspace | \thickspace}

\begin{document}

\preprint{LA-UR-05-8831}
\title{Construction and analysis of a simplified many-body neutrino model}

\author{Alexander Friedland} \email{friedland@lanl.gov}
\affiliation {Theoretical Division, T-8, MS B285,Los Alamos National Laboratory, Los Alamos, NM 87545, USA}

\author{Bruce H.J. McKellar} \email{b.mckellar@physics.unimelb.edu.au}
\affiliation{School of Physics, Research Center for High Energy
  Physics, The University of Melbourne, Victoria 3010, Australia}

\author{Ivona Okuniewicz} \email{ivona@physics.unimelb.edu.au}
\affiliation{School of Physics, Research Center for High Energy
  Physics, The University of Melbourne, Victoria 3010, Australia}

\date{February 1, 2006}

\begin{abstract}
  In dense neutrino systems, such as found in the early Universe, or
  near a supernova core, neutrino flavor evolution is affected by
  coherent neutrino-neutrino scattering. It has been recently
  suggested that many-particle quantum entanglement effects may play
  an essential role in these systems, potentially invalidating the
  traditional description in terms of a set of single-particle
  evolution equations. We model the neutrino system by a system of
  interacting spins, following an earlier work which showed that such
  a spin system can in some cases be solved exactly
  \cite{Friedland_&_Lunardini_1}. We extend this work by constructing
  an exact analytical solution to a more general spin system,
  including initial states with asymmetric spin distribution and,
  moreover, not necessarily aligned along the same axis.  Our solution
  exhibits a rich set of behaviors, including coherent oscillations
  and dephasing and a transition from the classical to quantum
  regimes. We argue that the classical evolution of the spin system
  captures the entire coherent behavior of the neutrino system, while
  the quantum effects in the spin system capture some, but not all, of
  the neutrino incoherent evolution.  By comparing the spin and
  neutrino systems, we find no evidence for the violation of the
  accepted one-body description, though the argument involves some
  subtleties not appreciated before. The analysis in this paper may
  apply to other two-state systems beyond the neutrino field.
\end{abstract}


\maketitle

\section{Introduction}

In many neutrino systems that are currently studied the rate of
incoherent interactions is low enough to be completely negligible, yet
\emph{coherent} interactions (refraction) play an important, even
crucial role. A classical example is provided by the case of solar
neutrinos: these neutrinos hardly scatter inside the Sun;
nevertheless, their coherent interactions with the solar matter plays
an essential role in their flavor evolution.  This is of course the
celebrated MSW theory of neutrino oscillations in matter
\cite{Wolfenstein, MS}.

The classical MSW theory describes neutrino propagation in a
background of normal matter (electrons, neutrons, protons). There
are systems, however, where the number density of neutrinos
themselves exceeds those of electrons and baryons, such as the
early universe, or the so-called hot bubble region in an exploding
supernova. Such systems additionally require a theory describing
neutrino self-refraction.

Early investigations treated the neutrino background analogously
to the MSW theory \cite{Fuller, Notzold}. Pantaleone
\cite{Pantaleone1992prd,Pantaleone1992plb}, McKellar and Thomson
\cite{McKellar_Thomson}, and Sigl and Raffelt
\cite{Sigl_Raffelt} showed, however, that the neutrino background is
distinguished by an important subtlety: the induced mass terms in
general have nonzero off-diagonal components in the flavor basis.
Physically, this means that flavor can be \emph{coherently}
exchanged between the neutrinos.

The authors of
\cite{McKellar_Thomson,Pantaleone1992prd,Pantaleone1992plb,Sigl_Raffelt}
constructed the equation of motion of the neutrino system by using a
one-body description for each neutrino. This treatment crucially
depends on the assumption that the state of the system can be
factorized into a product of one-particle states. If this is not the
case and the wavefunctions of individual neutrinos are entangled, a
very different treatment may be required.  {\it A priori}, it is not
obvious that neutrinos would not develop such entanglement, due to the
off-diagonal induced mass terms.  So the question arises whether this
entanglement exists and has a substantial impact on the coherent
flavor evolution of the neutrino ensemble. The answer could have a
significant impact on the predictions for the supernova neutrino
signal, the synthesis of heavy elements in the supernova, and possibly
Big Bang Nucleosynthesis and cosmology.

The question was recently examined by Friedland \& Lunardini (F\&L
I \cite{Friedland_&_Lunardini_2}, F\&L II
\cite{Friedland_&_Lunardini_1}) and by Bell, Rawlinson and Sawyer
(BRS \cite{Bell}). All three papers used a similar
setup, with a Hamiltonian that was restricted to ``forward
scattering'' only \cite{Pantaleone1992plb}. While F\&L I argued
that \emph{the coherent part} of the neutrino evolution should be
described by the one-particle formalism, the BRS paper reached an
opposite conclusion.

To make their argument, BRS considered the evolution of a system
initially in the flavor eigenstates. With this choice, the
one-particle formalism predicts no coherent flavor conversion and
thus conversion on the coherent time scale in this system would be
an indication of the breakdown of the one-particle description
(presumably through the formation of many-neutrino entangled
states). The numerical calculation performed by BRS seemed to
suggest the presence of such ``fast'' conversions, although the
calculations involved a relatively small numbers of neutrinos.

F\&L II \cite{Friedland_&_Lunardini_1} subsequently solved the
neutrino model introduced by BRS analytically for the special case of
equal numbers of each neutrino species and equal strength interactions
by mapping neutrino-neutrino interactions to spin-spin interactions.
The solution in the limit of many particles exhibited the
equilibration time that is precisely what would be expected from
incoherent scattering. The analytical solution thus supported the
one-particle description of the system.

One may wonder, however, if the initial state considered in F\&L
II, namely, equal numbers of spins ``up'' and ``down'' was somehow
special. Could entanglement appears with a more general setup?

In this paper we will present a generalization of the many-body
neutrino model introduced in F\&L II, in the hope of understanding
the quantum system better.  We again consider a system of many
neutrinos in which there are two flavor species, so that it maps
to a system of interacting spins (thanks to the $SU(2)$ symmetry
of the problem). We generalize the model to initial states in
which the species are not equally populated and where the initial
states are not necessarily in flavor eigenstates. We show that the
corresponding spin problem can still be solved exactly. The
resulting solution exhibits a rich set of behaviors, as will be
discussed in the following.

Even generalized in this way, the model still involves several
simplifications and it is important to spell these out.
\begin{itemize}
\item The model keeps only forward scattering terms. Thus, while the
  model should correctly capture coherent effects in a real neutrino
  system, conclusions about incoherent scattering effects in our model
  must be interpreted with care.
\item The momentum degrees of freedom are ignored. The effects of
  Fermi statistics are not included, that is the physical neutrino
  system is assumed to have a phase-space density much less than one.
  This is indeed satisfied everywhere outside the neutrinosphere in a
  supernova. The case of the neutrinos in the early universe may be
  more subtle \footnote{For a thermal neutrino distribution with zero
    chemical potential, the phase space density is given by
    $(\exp(E/kT)+1)^{-1}$. This means that while the neutrinos on the
    tail $(E\gg kT)$ are non-degenerate, those with $E\lesssim kT$ are
    mildly degenerate. In the later regime, the present approximation
    may be inadequate.  Of course, if the chemical potential is
    significant, the phase space density will be of order one in some
    regions.}.
\item The interaction strength between any two neutrinos is taken to
  be the same, ignoring the angular distribution of the neutrino
  momenta (see later). The model thus aims to describe the physical
  situation in an isotropic neutrino gas and hence may or may not
  capture all effects that could arise as a result of very anisotropic
  momentum distributions, such as those suggested in
  \cite{Sawyer2004,Sawyer2005}.
\end{itemize}
These limitations and assumptions should be kept in mind when relating
the results obtained for the spin system to the behavior of a real
neutrino gas.

The problem of the flavor evolution in dense neutrino systems
continues to receive a significant amount of attention. In addition to
the above mentioned papers
\cite{Friedland_&_Lunardini_2,Bell,Friedland_&_Lunardini_1,Sawyer2004,Sawyer2005},
the reader is referred to
\cite{Boyanovsky:2004xz,Sirera:2004rv,Strack:2005ux,Strack:2005jj,Ho:2005vj,DuanFuller2005}
for recent progress.


\section{Setup and Goals}
\label{sec:Hamiltonian}

\subsection{The Hamiltonian and Eigenvalues}

We follow \cite{Bell} and \cite{Friedland_&_Lunardini_1} by
considering a system consisting of interacting massless neutrinos
represented by plane waves in a box of volume $V$. Since our primary
motivation is to investigate coherent effects in the neutrino system,
in particular, the possible breakdown of the one-body approximation
due to flavor exchange, and not due to spatially dependent many-body
correlations, we focus on the ``forward'' neutral current interactions
between the neutrinos.  In other words, we drop the momentum degrees of
freedom and include only scattering events that preserve neutrino
momenta and those that \emph{exchange} the momenta
\cite{Pantaleone1992plb,Friedland_&_Lunardini_2},
\begin{eqnarray}
  \label{eq:forward}
  |\nu_x (\vec{k}) \nu_y(\vec{p})\rangle &\longrightarrow& |\nu_x
  (\vec{k}) \nu_y(\vec{p})\rangle,\\
  |\nu_x (\vec{k}) \nu_y(\vec{p})\rangle &\longrightarrow& |\nu_x
  (\vec{p}) \nu_y(\vec{k})\rangle.
\end{eqnarray}

In the ``usual" case of electrons, protons and neutrons in the
background, the waves scattered forward interfere coherently. For the
neutrino background, however, this is not necessarily so and our
model, in addition to coherent effects, captures some of the
incoherent effects as well
\cite{Friedland_&_Lunardini_2,Friedland_&_Lunardini_1}. The
identification of coherent and incoherent effects will be discussed at
length in what follows.

For $E_\nu \ll m_{Z^0}$ the neutral current interaction
Hamiltonian is

\begin{equation}
H_{int} = \frac{G_F}{\sqrt{2}}\left(\sum_\alpha \bar{\nu}_\alpha
\gamma^\mu\nu_\alpha\right)\left(\sum_\beta \bar{\nu}_\beta
\gamma_\mu\nu_\beta\right).
\end{equation}

Here the sum is over all $k$ flavors. The Hamiltonian is invariant
under a flavor $SU(k)$ symmetry. Let us consider only two neutrino
species, in which case the symmetry becomes $SU(2)$ and the flavor
space structure of the $\nu-\nu$ interaction becomes equivalent to
the interaction between pairs of spins. As explicitly shown in
F\&L II, the interaction energy of two neutrinos, 1 and 2, is
proportional to the square of the total angular momentum of the
corresponding spin system,
\begin{equation}
 \hat{L}^2 \equiv (\vec\sigma_1/2+\vec\sigma_2/2)^2 =
 (3/2+\vec\sigma_1\cdot\vec\sigma_2/2).
\end{equation}

The coefficient of proportionality dependents on the relative
angle between the neutrino momenta $\Theta$,
$g=\frac{\sqrt{2}G_F}{V}(1-\cos\Theta)$ \cite{Pantaleone1992plb}.
In a realistic neutrino system, the couplings are distributed
according to the distribution of the relative angles between
neutrino momenta. In order to make our system solvable, we will
simplify the problem and take all the couplings to be the same
\begin{equation}\label{eq:g_def}
    g=\frac{\sqrt{2}G_F}{V}.
\end{equation}
It is hoped that this simplification preserves the essential
features of the evolution \cite{Bell,Friedland_&_Lunardini_1}
(see, however, \cite{Sawyer2004,Sawyer2005}). Henceforth we study
this system of interacting spins to obtain information about the
neutrino system.

We will consider a system of $N+M$ spins, such that initially $N$
spins all have a certain orientation (for definiteness, without a loss
of generality, ``up") and the remaining $M$ spins all have a certain
different orientation. At $t=0$, thus, the $N$ spins combine in a
state of angular momentum $J_N=N/2$ and projection $m_N=N/2$, and the
M spins in a state of angular momenta $J_M=M/2$ and projection $m_M$.
In terms of the original neutrino system, we have a system of $N$
electron neutrinos, $|\nu_e\rangle$, and $M$ neutrinos in some other
state $|\nu_x\rangle$.  We give the answer for a general
$|\nu_x\rangle$ and then explicitly study two cases: $|\nu_x\rangle$
is flavor eigenstate $|\nu_\mu\rangle$ and $|\nu_x\rangle$ is a flavor
superposition state $(|\nu_e\rangle+|\nu_\mu\rangle)/\sqrt{2}$.

The Hamiltonian for this system is
\begin{widetext}
\begin{eqnarray}
\label{eq:H1}
 H = g \sum_{j=1}^{N+M-1}\sum_{k=j+1}^{N+M}
(\vec{\sigma}_j + \vec{\sigma}_k)^2  = g
\left(\sum_{j=1}^{N+M-1}\sum_{k=j+1}^{N+M} 2\vec{\sigma}_j \cdot
\vec{\sigma}_k + \frac{3}{4}(N+M)(N+M-1)\right).
\end{eqnarray}
\end{widetext}
which is related to the square of the total angular momentum of
the system \cite{Friedland_&_Lunardini_1},
\begin{eqnarray}
\label{eq:H2}
 \hat{J}^2&=&\left(
\sum_{j=1}^{N+M}\vec{\sigma}_j\right)^2
\nonumber\\
&=& \left(\sum_{j=1}^{N+M-1}\sum_{k=j+1}^{N+M} 2\vec{\sigma}_j
\cdot \vec{\sigma}_k + \frac{3}{4}(N+M)\right).
\end{eqnarray}
By comparing Eqs.~(\ref{eq:H1}) and (\ref{eq:H2}) we find
\begin{equation}
\label{eq:Hamiltonian} H_{int} = g[\hat{J}^2 +
\frac{3}{4}(M+N)(M+N-2)],
\end{equation}
with eigenvalues
\begin{eqnarray}
\label{eq:eigenvalues} E(J, N, M) =& g\left[J(J+1) +
\frac{3(M+N)(M+N-2)}{4}\right],
\end{eqnarray}
where
\begin{eqnarray}
 J_{\rm min} &\le& J \le J_{\rm max},\\
 \label{eq:Jmin}
    J_{\rm min}&=&{\rm max}(|m_N-m_M|,|N/2-M/2|),\\
 \label{eq:Jmax}
    J_{\rm max} &=& N/2+M/2.
\end{eqnarray}

\subsection{Goals}

We are interested in finding the probability, as a function of
time, $P_1(t)$, of one of the particles remaining in the ``spin
up" state if it was initially in the ``spin up" state. As
discussed in \cite{Friedland_&_Lunardini_1}, the time scale with
which this probability evolves, $t_{\rm eq}$, tells us whether the
evolution has coherent or incoherent nature. In particular, for
$M=N$,
\begin{eqnarray}
 \label{eq:t_coh}
   t_{\rm eq}^{coh} &\sim& (g N)^{-1}, \\
 \label{eq:t_inc}
   t_{\rm eq}^{inc} &\sim& (g \sqrt{N})^{-1},
\end{eqnarray}
for coherent and incoherent evolution correspondingly. In a large
spin system the coherent time scale is much shorter then the
incoherent one. One of our goals will be to see which timescales
are present in our solution under different initial conditions.

The second goal is to compare the coherent evolution we find to
the predictions of the one-particle formalism. According to this
formalism, the coherent evolution of a given neutrino is
determined by the following one-particle Hamiltonian
\cite{Pantaleone1992plb, Pantaleone1992prd}:
\begin{eqnarray}
\label{eq:H_oneparticle}
 H_{1-part}=\frac{\sqrt{2}G_F}{V}\sum_i |\psi_i\rangle\langle
 \psi_i|.
\end{eqnarray}
Here, $\psi_i$ is the flavor state of the $i$th ``background''
neutrino i.e. the background is all the neutrinos except for the one
for which the equation is written.  Explicitly, for two neutrino
species ($\nu_e$ and $\nu_\mu$)  $|\psi_i\rangle =
\left[\begin{array}{c} \nu_e^i \\ \nu_\mu^i\end{array}\right]$.  The
sum runs over all ``background" neutrinos.


\section{The Probability of Spin Preservation}

\subsection{Result}

As we show in this Section, the evolution of our system can be
solved exactly. In the interests of clarity, we begin by
displaying the answer for $P_1(t)$:
\begin{eqnarray}
\label{eq:form of probability}
 P_1(t)&=&\frac{1}{2}+\sum_{J=J_{\rm
min}}^{J_{\rm max}} C(J)\nonumber\\
 &+& \sum_{J=J_{\rm min}}^{J_{\rm max}-1} \eta(J)\ \cos[gt(2J+2)].
\end{eqnarray}

The first two terms in Eq.~(\ref{eq:form of probability}) give the
mean value of the probability: $P_1=1/2$ means complete
depolarization, or in the language of the neutrino system, an
equal incoherent mixture of the two flavors; the second term thus
given the degree of polarization of the mean (``equilibrated")
state. The last term contains the time evolution of the system.

The limits of the summation $J_{\rm min}$ and $J_{\rm max}$ are given
in Eq.~(\ref{eq:Jmin}) and (\ref{eq:Jmax}). The coefficients $C(J)$
and $\eta(J)$ given by
\begin{widetext}
\begin{eqnarray}
\label{eq:CJ}
 C(J) &=& (-1)^{2J-(m_N+m_M)+(N+M)/2+1}(1+2 J)
\frac{1}{N} \sqrt{(N/2)(N/2+1)(N+1)}
\nonumber\\
&\times&
|\langle N/2, m_N, M/2,m_M |J, m_N+m_M\rangle|^2\nonumber\\
&\times&
 \begin{pmatrix}
    J       & 1 & J\\
    -(m_N+m_M) & 0 & m_N+m_M
  \end{pmatrix}
  \left\{
  \begin{array}{ccc}
    N/2 & J  & M/2 \\
    J   & N/2 & 1
  \end{array}
  \right\},\\
\label{eq:etaJ}
 \eta(J)&=&  (-1)^{2J-(m_N+m_M)+(N+M)/2} \sqrt{(1+2 J)(3+2 J)}
\frac{2}{N}\sqrt{(N/2)(N/2+1)(N+1)}\nonumber\\
&\times&
\langle N/2, m_N, M/2,m_M|J, m_N+m_M\rangle
\langle N/2, m_N, M/2,m_M|J+1, m_N+m_M\rangle\nonumber\\
&\times&
 \begin{pmatrix}
    J  + 1     & 1 & J\\
    -(m_N+m_M) & 0 & m_N+m_M
  \end{pmatrix}
 \left\{
  \begin{array}{ccc}
    N/2 & J + 1 & M/2 \\
    J   & N/2 & 1
  \end{array}
  \right\}.
\label{eq:C}
\end{eqnarray}
\end{widetext}

The inner products in the last two equations are Clebsch-Gordan
coefficients, the objects in the parentheses are $3j$-coefficients
and those in curly brackets are $6j$-coefficients. For
definitions, see, {\it e.g.}, \cite{Landavshitz}.

The rest of this Section presents two complementary derivations of
these results. The derivations are somewhat technical and the reader
primarily interested in the analysis of the rich physical properties
of the solution may wish to skip to Sect.~\ref{sec:superposition_bg}
and return to this Section later, as needed.

\subsection{Construction of the probability: overview}

This solution can be found in either of two ways, both of which
provide important, complementary physical insights into the spin
system. These insights will prove very useful later, as we discuss
the physical properties of the solution. Additionally, one or the
other method may be useful for addressing still more general spin
configurations. Correspondingly, we show both methods.
\begin{itemize}
  \item The first approach is to ``split off" the first spin from
  the system, so that the remaining $N+M-1$ spins forms ``a background" it
  interacts with. The solution is constructed by first coupling the angular
  momenta of the remaining spins and then coupling the first spin to the
  result. This method generalizes the idea employed in
  \cite{Friedland_&_Lunardini_1}, but without relying on the
  symmetries specific to the $N=M$ case.
  \item The second approach is to treat the first spin as a
  part of the system. We solve for the evolution of all spins that
  start out in the ``up" state. The solution can be found by
  observing that even with $N\ne M$ the system possesses a very
  high degree of symmetry: all spins that start out in the same
  state evolve in the same way. This means, as we will show, that the problem
  can be reduced to that of just two coupled angular momenta.
\end{itemize}

\subsection{Method I: splitting off the first spin}

The outline of this approach is as follows. The time evolution of
the system is easily written down in the basis of \emph{total
angular momentum}, since, as explained in
Sect.~\ref{sec:Hamiltonian}, this is the Hamiltonian eigenbasis.
Correspondingly, we begin in Sect.~\ref{sec:density matrix} by
constructing the density matrix for the system in this basis. In
Sect.~\ref{sect:split_spin}, this density matrix is rotated to a
basis in which the first spin has a well-defined value. Finally,
in Sect.~\ref{sec:split_spin_P1} the probability $P_1(t)$ is
found.


\subsubsection{Constructing the many-body density matrix in the
total angular momentum basis} \label{sec:density matrix}


Our system begins (at time $t=0$) in the state
\begin{equation}
\label{eq:initial state} S(0)=\lvertical j_N,
m_N\rightangle\otimes\lvertical j_M,m_M\rightangle.
\end{equation}
Recall $j_N=N/2$ is the total angular momentum of all the spin up
particles, each with angular momentum $\half$, and  projection
along the $\hat{z}$ direction, $m_N$. Also, $j_M=M/2$ is the
angular momentum of all the background particles, each with
angular momentum $\half$ and projection, $m_M$. Rotating the
initial state, Eq.(\ref{eq:initial state}), to the total angular
momentum ($J$) basis and evolving it to time $t$, we have,
\begin{eqnarray}
\label{eq:state original basis} S(t) = \sum_{J=J_{\rm min}}^{J_{\rm
max}} &&\leftangle j_N, j_M, J, m\midle j_N, j_M, m_N, m_M
\rightangle\nonumber\\
 &&\times e^{-itE(J, N, M)}\midle j_N, j_M,
J,m \rightangle.
\end{eqnarray}
Here $\leftangle j_N, j_M, J, m\midle j_N, j_M, m_N, m_M
\rightangle$ is the Clebsch-Gordan coefficient where $j_N$ and $j_M$
are coupled to the total angular momentum of the system $J$ with
projection in the $\hat{z}$ direction of $m=m_N+m_M$.

The density matrix is defined as
\begin{equation}
\rho(t) = \lvertical S(t) \rightangle\leftangle S(t) \rvertical.
\end{equation}
Hence we have,
\begin{widetext}
\begin{eqnarray}
\label{eq:density matrix} \nonumber \rho(t)&=\sum_{J=J_{\rm
min}}^{J_{\rm max}}\sum_{J'=J_{\rm min}}^{J_{\rm max}}&e^{-it\Delta
E(J,J')}\leftangle j_N, j_M, J, m\midle j_N, j_M, m_N, m_M
\rightangle\leftangle j_N, j_M, J', m\midle j_N, j_M, m_N, m_M \rightangle \\
&&\midle j_N, j_M, J,m\rightangle\leftangle j_N, j_M, J',m\rvertical.
\end{eqnarray}
 \end{widetext}
Here $\Delta E(J,J')= g[J(J+1) - J'(J'+1)]$ is the difference between
the eigenvalues $E(J,N,M)$ and $E(J',N,M)$.


\subsubsection{Construction of the density matrix in the basis
$|\half, \lambda\rangle\otimes |j, \mu\rangle$}
\label{sect:split_spin}

\begin{figure}[bt]
\begin{center}
\fbox{\epsfig{file=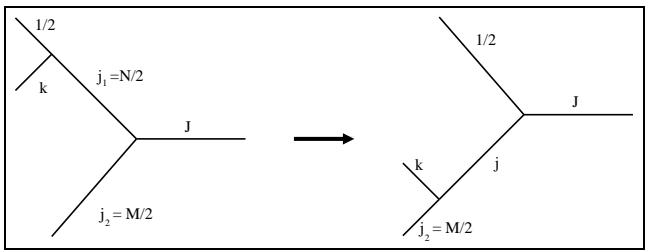, height=3cm, angle=0}}
\end{center}
\caption{Graphical representation of recoupling.  The diagram on the
left
represents the original coupling: the net angular momentum ($j_1$)
of the $N$ spins pointing up and the net angular momentum ($j_2$) of
the remaining $M$ spins are coupled to the total angular momentum of
the system ($J$).  The diagram on the right
represents the new coupling (see main text for description). $k$ is
the momentum of $N-1$ ``up" spins.} \label{fig:recoupling}
\end{figure}

To find the probability of spin preservation it is convenient to
again change the basis, this time to  $|\half,
\lambda\rangle\otimes |j, \mu\rangle$.  $\quad |\half,
\lambda\rangle$ is the state of one of the spin-1/2 particles with
projection $\lambda$ and $|j, \mu\rangle$ is the state of $M+N-1$
remaining particles with angular momenta j and projection $\mu$.
The density matrix in the new basis is constructed in this
subsection and then the probability is found in the next
subsection. We start with the state of  Eq. (\ref{eq:state
original basis}),
and transform it to the new basis. We will omit the limits on the
summation signs for the next few equations but will comment on
these later.

To change to the preferred basis ($|\half, \lambda\rangle\otimes |j,
\mu\rangle$) we modify  how the angular momenta couple to the total
angular momentum, $J$.  This can be done in any way that is
convenient. We for the moment ``remove" the first spin up and
combine the $N-1$ spins up (with net angular momentum $k=j_N-1/2$)
and the remaining $M$ spins (with net angular momentum $j_M$) into
an object with angular momentum $j$. We then couple the removed spin
and $j$ to the total $J$. Therefore $j$ is the result of coupling
$j_M$ and $k$.  See Fig.~\ref{fig:recoupling} for a pictorial
interpretation.

The state $ \lvertical j_N, j_M, J, m\rightangle$ is represented in
the new basis as
\begin{eqnarray}
\lvertical j_N, j_M, J, m\rightangle &=&
\sum_j\leftangle \frac{1}{2}, (k\thickspace j_M)j; J \midle(\frac{1}{2}\thickspace  k)j_N, j_M; J \rightangle
\nonumber\\
&\times&\midle \frac{1}{2} , (k\thickspace j_M)j; J, m\rightangle.
\end{eqnarray}

The notation $(a\thickspace b)c$ indicates that $a$ and $b$ couple
to $c$. Here the recoupling coefficient $\leftangle \frac{1}{2},
(k\thickspace j_M)j; J \midle(\frac{1}{2}\thickspace  k)j_N, j_M; J
\rightangle$ is proportional to the 6-j coefficient,
\begin{widetext}
\begin{equation}
 \leftangle \frac{1}{2}, (k\thickspace j_M)j; J \midle(\frac{1}{2}\thickspace  k)j_N, j_M; J \rightangle\equiv (-1)^{\half + k +J + j_M}\left[(2j_N + 1)(2j +1)\right]^\half\left\{\begin{array}{ccc}
\frac{1}{2} & k & j_N \\
j_M & J & j\\
\end{array}\right\}.
\end{equation}
\end{widetext}
For a definition and explanation of 6-j coefficients see \cite{Brink} or
\cite{Landavshitz}, Eq.~(108.6).
The right side of this coefficient represents the original
coupling (the left side of Fig.~\ref{fig:recoupling}). The left
side represents the changed coupling (the right side of
Fig.~\ref{fig:recoupling}). The 6-j coefficient is a consequence
of changing the coupling.

Rotating from the basis $\midle \frac{1}{2} , (k\thickspace j_M)j;
J, m\rightangle$ to the basis $\lvertical \frac{1}{2}, \lambda
\rightangle\otimes\lvertical j, \mu\rightangle$ and substituting
$k=j_N-1/2$, we have,
\begin{widetext}
\begin{equation}
\begin{split}
\label{eq:recoupled state}
\lvertical j_N, j_M, J, m\rightangle =&\sum_{j,\lambda, \mu}
(-1)^{J + j_M+j_N}\left[(2j_N + 1)(2j +1)\right]^\half
\leftangle\frac{1}{2} , j, \lambda, \mu \midle \frac{1}{2}, j; J, m \rightangle
\\&\times\left\{\begin{array}{ccc}
\half & j_N-\half & j_N \\
j_M & J & j\\
\end{array}\right\}
\midle \frac{1}{2}, \lambda \rightangle\otimes \midle j, \mu
\rightangle.
\end{split}
\end{equation}
\end{widetext}

Finally, substituting Eq.~(\ref{eq:recoupled state}) into
Eq.~(\ref{eq:state original basis}) yields
\begin{widetext}
\begin{eqnarray}
\label{eq:final state}
 S(t) &=&\sum_{J, j,\lambda, \mu}
(-1)^{J + j_M + j_N}\left[(2j_N + 1)(2j +1)\right]^\half
 e^{-itE(J,N,M)}\leftangle j_N, j_M, J, m \midle  j_N,
j_M, m_N, m_M \rightangle
 \nonumber\\
&\times&\leftangle\frac{1}{2} , j, \lambda, \mu \midle
\frac{1}{2}, j; J, m \rightangle \left\{\begin{array}{ccc}
\half & j_N-\half & j_N \\
j_M & J & j\\
\end{array}\right\}
\midle \frac{1}{2}, \lambda \rightangle\otimes \midle j, \mu
 \rightangle.
\end{eqnarray}
\end{widetext}

Recall that the first $N$ spins are initially in a state with total
angular momentum $j_N=N/2$. The initial state of the test spin
is consequently the same as that of  the other $N-1$
particles, ``up'' ($\lambda=+1/2$).  

It is now simple to construct the density matrix in this new basis
using Eq.~(\ref{eq:final state}) and $\rho =  \lvertical
S(t)\rightangle \leftangle S(t)\rvertical$.  We present the density
matrix in component form,
\begin{widetext}
\begin{equation}
\begin{split}
\label{eq:density matrix new basis}
 \rho_{\half\thickspace\lambda,\thickspace j\thickspace\mu\thickspace;\thickspace\half\thickspace\lambda',\thickspace j'\thickspace\mu'}(t)=
   \sum_{J,J'}
 &e^{-it\Delta E(J,J')}(-1)^{J-J'}(2j_N + 1)(2j +1)
 \leftangle j_N, j_M, J, m \midle  j_N, j_M, m_N, m_M \rightangle\\
 &\times\leftangle j_N, j_M, J', m \midle  j_N, j_M, m_N, m_M \rightangle
 \leftangle\frac{1}{2} , j, \lambda, \mu \midle \frac{1}{2},
j; J, m \rightangle \\
&\times\leftangle\frac{1}{2} , j', \lambda', \mu' \midle
\frac{1}{2}, j'; J', m \rightangle\left\{\begin{array}{ccc}
\half & j_N-\half & j_N \\
j_M & J & j\\
\end{array}\right\}
\left\{\begin{array}{ccc}
\half & j_N-\half & j_N \\
j_M & J' & j'\\
\end{array}\right\}.
\end{split}
\end{equation}
\end{widetext}


\subsubsection{Probability of Spin Preservation}
\label{sec:split_spin_P1}

The probability that the first spin remains in the up state can be
found from the density matrix. In general the probability of an
eigenvalue $a_i$ represented by the operator A is
$\mbox{Tr}(A\rho)$.  In this case the operator is diagonal and hence
probabilities are just the diagonal components of the density
matrix.  These components of the density matrix which give
probabilities are those where $j=j'$, $\mu=\mu'$, and $\lambda =
\lambda'$, {\it i.e.}, $\rho_{\half\thickspace\lambda,\thickspace
j\thickspace\mu;\half\thickspace\lambda,\thickspace
j\thickspace\mu}$.  Furthermore we are looking for the probability
of the first spin remaining in the spin up state so that $\lambda =
+\half$. Using this information together with Eq. (\ref{eq:density
matrix new basis}) we find the probability to be
\begin{widetext}
\begin{eqnarray}
\label{eq:probability} P_1(t) &=& \sum_{j, \mu}
\rho_{\half\thickspace\half,\thickspace
j\thickspace\mu;\half\thickspace\half,\thickspace
j\thickspace\mu}=\sum_{J,J',j, \mu}
 e^{-it\Delta E(J,J')}(-1)^{J-J'}(2j_N + 1)(2j +1)
 \leftangle j_N, j_M, J, m \midle  j_N, j_M, m_N, m_M
 \rightangle \nonumber\\
&\times&\leftangle j_N, j_M, J', m \midle  j_N, j_M, m_N, m_M
\rightangle
 \leftangle\frac{1}{2} , j, \half, \mu \midle
\frac{1}{2}, j; J, m \rightangle
 \leftangle\frac{1}{2} , j, \half,
\mu' \midle \frac{1}{2}, j; J', m \rightangle \nonumber\\
&\times&\left\{\begin{array}{ccc}
\half & j_N-\half & j_N \\
j_M & J & j\\
\end{array}\right\}
\left\{\begin{array}{ccc}
\half & j_N-\half & j_N \\
j_M & J' & j\\
\end{array}\right\}.
\end{eqnarray}
\end{widetext}

The third Clebsch-Gordan coefficient in Eq. (\ref{eq:probability}),
$\leftangle\frac{1}{2} , j, \half, \mu \midle \frac{1}{2}, j; J, m
\rightangle$, represents the  coupling of $\half$ and $j$ to $J$.
This gives $J-\half \leqslant j \leqslant J+\half$.  In this same
Clebsch-Gordan notice that we must have $\half + \mu = m$, so that
the summation over $\mu$ is unnecessary.  Analyzing the fourth
Clebsch-Gordan coefficient, $\leftangle\frac{1}{2} , j, \half, \mu'
\midle \frac{1}{2}, j; J', m \rightangle$, further simplifies the
equation.  This coefficient shows that we couple $\half$ and $j$ to
$J'$.  Hence we have,  $|j-\half| \leqslant J' \leqslant j+\half$.
Now if $j=J+\half$ then  $J \leqslant J' \leqslant J+ 1$, and if
$j=J-\half$ then $|J-1| \leqslant J' \leqslant J$. Therefore $J'=
|J-1|, J, J+1$.

Summing over $J'$ we find
\begin{equation}
\begin{split}
\label{eq:final probability}
 P_1(t) &=
\sum_{J=J_{\rm min}}^{J_{\rm max}} \tilde{C}(J)
\\     &+\sum_{J=J_{\rm min}}^{J_{\rm max}-1}\eta(J)\cos[t\Delta
E(J+1,J)],
\end{split}
\end{equation}
where
\begin{equation}
\label{eq:constant}
\begin{split}
\tilde{C}(J)&= \sum_{j=J-\half}^{J+\half}(2j_N +1)(2j+1) \\
 &\times\left|\leftangle j_N, j_M, J, m
\midle j_N, j_M, m_N, m_M \rightangle
\right|^2\\&\times\left|\leftangle\frac{1}{2} , j, \half, m-\half
\midle \frac{1}{2}, j; J, m \rightangle \right|^2
       \\&\times\left| \left\{\begin{array}{ccc}
\half & j_N-\half & j_N \\
j_M & J & j\\
\end{array}\right\}\right|^2
\end{split}
\end{equation}
and
\begin{equation}
\label{eq:eta}
\begin{split}
\eta(J) &= -2(2j_N +1)(2J+2)
\\&\times\leftangle j_N, j_M, J, m \midle  j_N, j_M, m_N, m_M \rightangle
\\&\times\leftangle j_N, j_M, J+1, m \midle  j_N, j_M, m_N, m_M \rightangle
\\&\times    \leftangle\frac{1}{2} , J+\half, \half, m-\half \midle \frac{1}{2}, J+\half; J, m \rightangle
\\&\times\leftangle\frac{1}{2} , J+\half, \half, m-\half \midle \frac{1}{2}, J+\half ; J+1, m \rightangle
\\&\times \left\{\begin{array}{ccc}
          \half & j_N-\half & j_N \\
           j_M & J & J+\half\\
          \end{array}\right\}
          \left\{\begin{array}{ccc}
          \half & j_N-\half & j_N \\
          j_M & J+1 & J+\half\\
          \end{array}\right\}.
\end{split}
\end{equation}
Although it is not immediately obvious, the answer just obtained
coincides with Eqs.~(\ref{eq:form of probability}), (\ref{eq:CJ}),
and (\ref{eq:etaJ}). Notice that $\tilde{C}(J)$ is not the same as
$C(J)$ in Eq.~(\ref{eq:CJ}) since, unlike Eq.~(\ref{eq:form of
probability}), Eq.~(\ref{eq:final probability}) does not contain the
1/2 term.


\subsection{Method II: symmetry of the entire system}
\label{sect:methodII}

The second method is based on an observation that since the
coupling strength is the same for all spins the system possesses a
very high degree of symmetry. In particular, all spins that start
out in the same state evolve in the same way (up to relabeling).
This simple observation proves to be very powerful for our
analysis: it means that \emph{all spins that start out in the
``up'' state combine in a single composite object with the angular
momentum $L_1=N/2$ at any moment in time, not just at $t=0$}.  The
same can be said about the remaining $M$ spins ($L_2=M/2$). Thus
the problem reduces to a textbook case of just two interacting
angular momenta.

More precisely, at any moment in time, the system is described as
a superposition of states $i$ with definite values of total
angular momentum, $J_i$. Each such state is obtained as a result
of symmetrizations and antisymmetrizations of individual spin
wavefunctions. The highest $J$ state is a completely symmetrical
combination, the next one is obtained by performing one
antisymmetrization, etc (the general rules are given by Young's
tableaux). The important point is that by symmetry all spins that
started out in the same state can only be symmetrized.  That means
that all spins that started out in the same state can be assembled
in a single object with the total (unchanging in time!) angular
momentum. This argument holds for any $t$.

It is conceptually straightforward to compute the expectation value of
the operator $j_z^{(1)}$ (the $z$ component of the angular momentum of
the first composite object) at any moment in time. The expectation
value of the corresponding operator for each spin, $s_z^{(1)}$, can
then be found by simply dividing by $N$, by symmetry.

Let us denote the state of the whole system by $|S(t)\rangle$. As
before, the system evolves according to
\begin{equation}
  \label{eq:gen_evol}
  |S(t)\rangle = \sum_{J=J_{\rm min}}^{J_{\rm max}} |J\rangle\langle
  J| S(0)\rangle e^{-i t E_J},
\end{equation}
where $J_{min}$ and $J_{max}$ are given in Eqs.~(\ref{eq:Jmin})
and (\ref{eq:Jmax}), $E_J\equiv E(J,N,M))$ given in
Eq.~(\ref{eq:eigenvalues}) (the irrelevant constant piece can be
dropped).

We need to find the probability of the first spin being in the
``up'' state, $P_1(t)=\langle S(t)|+1/2\rangle_1\langle
+1/2|_1|S(t)\rangle$. It follows from the definition of
  $s_z^{(1)}=1/2(|+1/2\rangle\langle +1/2|-|-1/2\rangle\langle -1/2|)$
  and the completeness relation $(|+1/2\rangle\langle
  +1/2|+|-1/2\rangle\langle -1/2|=1)$ that this probability is
related to the expectation value of the angular momentum of that
spin,
\begin{equation}\label{eq:P1_def}
     P_1(t)=\langle S(t)|s_z^{(1)}|S(t)\rangle+1/2.
\end{equation}
Let us now compute $\langle S(t)|s_z^{(1)}|S(t)\rangle$:
\begin{widetext}
\begin{eqnarray}
 \label{eq:doublesum}
  \langle S(t)|s_z^{(1)}|S(t)\rangle =
  \frac{1}{N} \langle S(t)|j_z^{(1)}|S(t)\rangle =
  \frac{1}{N} \sum_{J=J_{\rm min}}^{J_{\rm max}}
  \sum_{J'=J_{\rm min}}^{J_{\rm max}}\langle S(0)|J'\rangle\langle
  J'|j_z^{(1)}|J\rangle\langle J| S(0)\rangle e^{i t
  (E_J-E_{J'})}.
\end{eqnarray}
\end{widetext}
The products $\langle J| S(0)\rangle$, $\langle S(0)|J'\rangle$ are
just the Clebsch-Gordan coefficients $\langle N/2, m_N, M/2,m_M,| J,
m_N+m_M\rangle$, $\langle N/2, m_N, M/2,m_M,|J',m_N+m_M\rangle$
(written so because they are real). The problem of finding the
expectation value $\langle J'|j_z^{(1)}|J\rangle$ is solved, {\it
  e.g.}, in \cite{Landavshitz}. Using Eqs. (109.2), (109.3), (29.13)
there, and the fact that the operator $j_z^{(1)}$ is a $q=0$ component
of a $k=1$ spherical tensor (see (107.1) of \cite{Landavshitz} for the
exact definition), we get
\begin{eqnarray}
  \label{eq:matrix_element}
  \langle J'|j_z^{(1)}|J\rangle &=&
 \begin{pmatrix}
    J'       & 1 & J\\
    -(m_N+m_M) & 0 & m_N+m_M
  \end{pmatrix}\nonumber\\
&\times& \sqrt{(1+2 J)(1+2 J')}  \left\{
  \begin{array}{ccc}
    N/2 & J'  & M/2 \\
    J   & N/2 & 1
  \end{array}
  \right\}\nonumber\\
&\times&
\sqrt{(N/2)(N/2+1)(N+1)}\nonumber\\
&\times& (-1)^{q},
\end{eqnarray}
where
\begin{eqnarray}\label{eq:def_q}
    q&\equiv& max[J,J']-(m_N+m_N)+N/2+M/2\nonumber\\
    &+&min[J,J']+1.
\end{eqnarray}

The first line in Eq.~(\ref{eq:matrix_element}) expresses the
dependence of the matrix element on the $z$-component of the
angular momentum of the whole system (Eq. (109.2) of
\cite{Landavshitz}). The second line contains the dependence of
the reduced matrix element on the total angular momenta $J$ and
$J'$ (Eq. (109.3) of \cite{Landavshitz}). The third line is the
reduced matrix element of the angular momentum of the first
composite object, $\langle j^{(1)}|j^{(1)}|
j^{(1)}\rangle=\sqrt{L_1(L_1+1)(1+2L_1)}$. Finally, the last line
contains the sign factor collected from all the ingredients.

The 6j-symbol enforces the following selection rule:
\begin{equation}
  \label{eq:sel_rule}
  J'=J,J\pm 1.
\end{equation}
The double sum in Eq.~(\ref{eq:doublesum}) then reduces to
\begin{eqnarray}
  \label{eq:gen_answer}
\lefteqn{\langle S(t)|s_z^{(1)}|S(t)\rangle =}\nonumber\\
 &&\frac{1}{N} \sum_{J=J_{\rm min}}^{J_{\rm max}}
|\langle N/2, m_N, M/2,m_M,|J, m_N+m_M\rangle|^2\nonumber\\
&\times&
 \begin{pmatrix}
    J       & 1 & J\\
    -(m_N+m_M) & 0 & m_N+m_M
  \end{pmatrix}\nonumber\\
&\times& (1+2 J)  \left\{
  \begin{array}{ccc}
    N/2 & J  & M/2 \\
    J   & N/2 & 1
  \end{array}
  \right\}\nonumber\\
&\times&
\sqrt{(N/2)(N/2+1)(N+1)}\nonumber\\
&\times&
(-1)^{2J-(m_N+m_N)+N/2+M/2+1}\nonumber\\
&+&  \frac{1}{N} \sum_{J=J_{\rm min}}^{J_{\rm max}-1}
\langle N/2, m_N, M/2,m_M,|J, m_N+m_M\rangle\nonumber\\
&\times&
\langle N/2, m_N, M/2,m_M,|J+1, m_N+m_M\rangle\nonumber\\
&\times&
 \begin{pmatrix}
    J  + 1     & 1 & J\\
    -(m_N+m_M) & 0 & m_N+m_M
  \end{pmatrix}\nonumber\\
&\times& \sqrt{(1+2 J)(3+2 J)}  \left\{
  \begin{array}{ccc}
    N/2 & J + 1 & M/2 \\
    J   & N/2 & 1
  \end{array}
  \right\}\nonumber\\
&\times&
\sqrt{(N/2)(N/2+1)(N+1)}\nonumber\\
&\times&
(-1)^{2J-(m_N+m_N)+N/2+M/2}\nonumber\\
&\times&2\cos [2gt(J+1)],
\end{eqnarray}
which is the answer given earlier in Eqs.~(\ref{eq:form of
probability}), (\ref{eq:CJ}), and (\ref{eq:etaJ}).


\section{Analysis: a flavor superposition background}
\label{sec:superposition_bg}

As already mentioned, the solution in Eqs.~(\ref{eq:form of
  probability}), (\ref{eq:CJ}), and (\ref{eq:etaJ}) contains very rich
physics found in the actual neutrino system. First, let us show
that it contains both coherent oscillations and incoherent
equilibration (decay). For that, let us consider the case
\begin{equation}
\label{eq:quantum numbers_super}
 m_N=\frac{N}{2}, \hskip0.3truecm  m_M =0.
\end{equation}
In the neutrino system, this corresponds to the first $N$ neutrinos
initially being in the $|\nu_e\rangle$ flavor eigenstate, while the
remaining $M$ neutrinos starting out in the maximally mixed state.  In
this setup, the one-particle formalism predicts coherent evolution.

\begin{figure}[htbp]
  \centering
  \includegraphics[width=0.47\textwidth]{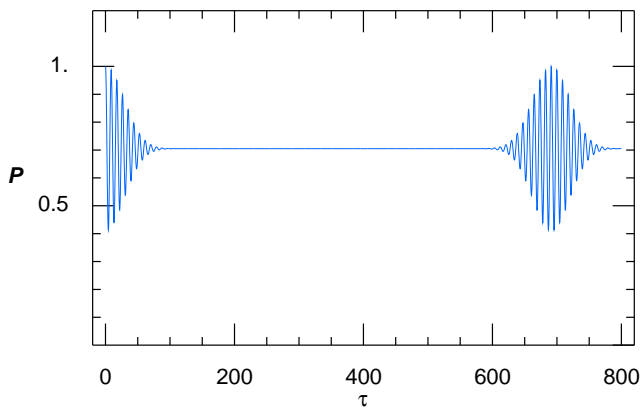}
  \caption{Time evolution of $P_1(t)$ in the case  $N=100$, $M=120$,
$m_N=N/2$, $m_M=0$. The time variable is defined as $\tau\equiv g t
(M+N)$.}
  \label{fig:coherent}
\end{figure}

Indeed, our solution confirms this. The explicit form of the
coefficients $C(J)$ and $\eta(J)$ is given in
Eqs.~(\ref{eq:C2_superposition}) and
(\ref{eq:eta2_superposition}). Using these values, we plot the
behavior of the expectation value $P_1(t)$ for $N=100$, $M=120$ in
Fig.~\ref{fig:coherent}.  We see that the evolution exhibits both
oscillations and decay. The oscillations reflect coherent
behavior, while the decay occurs on a longer time scale, as would
be expected for incoherent evolution.

This assertion can be made quantitative.  The key test of
coherence/incoherence of the evolution is the scaling of the evolution
times with the number of particles.  In the limit of large $N$ and
$M$, it is possible to show that a single ``wavetrain'' in
Fig.~\ref{fig:coherent} (oscillations plus decay) is described
analytically by the following formula:
\begin{eqnarray}
  \label{eq:coh_analyt}
  P_1(t) &=& \frac{1}{2} \left(1+\frac{N^2}{N^2+M^2} \right)\nonumber\\
  &+&
  \frac{1}{2} \left(1-\frac{N^2}{N^2+M^2} \right)
  \exp\left[-\frac{g^2 t^2}{\frac{2}{N}+2 \frac{N}{M^2}} \right]\nonumber\\
  &\times&\cos\left(g\sqrt{N^2+M^2}t \right).
\end{eqnarray}

A detailed derivation of Eq.~(\ref{eq:coh_analyt}) is given in
Appendix \ref{sec:appendix_coh}. Here, we notice that the
oscillation period scales as $\propto (N^2+M^2)^{-1/2}$, as
expected for coherent evolution, while the decay time goes like
$N^{-1/2}(2+2 N^2/M^2)^{1/2}$, indicating its incoherent nature
({\it cf} Eqs.~(\ref{eq:t_coh}), (\ref{eq:t_inc})).

We further observe that the frequency of the coherent oscillations
agrees with the predictions of the one-particle formalism for a
neutrino in the background of $M$ maximally neutrinos
$(|\nu_e\rangle+|\nu_\mu\rangle)/\sqrt{2}$ and $N-1 \simeq N$
neutrinos in the $|\nu_e\rangle$ state. Indeed, the one-particle
oscillation Hamiltonian (\ref{eq:H_oneparticle}) in this case is
\begin{eqnarray}
 H_{1-part}=\frac{\sqrt{2}G_F}{V}\sum_i |\psi_i\rangle\langle \psi_i|
 = g\begin{pmatrix}
    N+\frac{M}{2} & \frac{M}{2} \\
    \frac{M}{2} & \frac{M}{2}
  \end{pmatrix},
\end{eqnarray}
so that the oscillation frequency, given by the difference of the
eigenvalues, is precisely $g\sqrt{N^2+M^2}$.

A very valuable physical insight can be gained from the idea that
underlies Method II of solving for the evolution of the system,
namely, that all spins in the same initial state always combine to
form an object with a certain definite value of the angular momentum.
As explained in Sect.~(\ref{sect:methodII}), this means that the
system can be reduced to just two coupled angular momenta, $j_N=N/2$
and $j_M=M/2$. When the numbers of spins $N$ and $M$ are sufficiently
large, the two composite angular momenta behave as nearly classical
objects. They precess about the direction of the total angular
momentum of the system. These are the fast oscillations seen in
Fig.~\ref{fig:coherent}.

Quantum-mechanically, a system that has a definite value of the
total angular momentum does not simultaneously have definite
projections of the individual angular momentum vectors that comprise
it. Correspondingly, after a while the components of the angular
momenta that are transverse to the total angular momentum undergo
``quantum wash-out". The system equilibrates to a state in which only
the components of the constituent angular momenta along the
direction of the total angular momentum remain.

Let us check this quantitatively. The total angular momentum of the
system (in the classical limit) makes an angle
$\cos\alpha=N/\sqrt{N^2+M^2}$ with the positive $z$ direction. Hence,
the projection of the net angular momentum of the first $N$ spins on
the direction of the total angular momentum of the whole system is
$N/2\times\cos\alpha$. After a sufficient amount of time this is the
only component that remains, the transverse components are
washed-out. Projecting it back on the $z$-axis, we get
$N/2\times\cos^2\alpha$ or, using Eqs.~(\ref{eq:P1_def}),
(\ref{eq:doublesum}),
\begin{equation}
    \bar{P}_1=\frac{1}{2}+\frac{1}{N}\frac{N}{2}\frac{N^2}{N^2+M^2}
    = \frac{1}{2}\left(1+\frac{N^2}{N^2+M^2}\right),
\end{equation}
precisely in agreement with Eq.~(\ref{eq:coh_analyt}).

Additional insight about classical and quantum features of the
evolution can be gained by restoring the factors of the Plank's
constant in Eq. (\ref{eq:coh_analyt}). It is simple to see that the
product $g t$ comes with a one factor of $\hbar$. The logic is as
follows: (i) from Eq. (\ref{eq:eigenvalues}) it is obvious that the
energy is proportional to $\hbar^2$ (from the angular momentum
squared factor); (ii) next, in computing the evolution phase
($\exp(i E t/\hbar)$), we divide by one power of $\hbar$; (iii) this
leads to one power of $\hbar$ in the argument of the cosine in  Eq.
(14), {\it i.e.}, $gt$ comes with one factor of $\hbar$. Eq.
(\ref{eq:coh_analyt}) then reads
\begin{eqnarray}
  \label{eq:coh_analyt_hbar}
  P_1(t) &=& \frac{1}{2} \left(1+\frac{N^2}{N^2+M^2} \right)\nonumber\\
  &+&
  \frac{1}{2} \left(1-\frac{N^2}{N^2+M^2} \right)
  \exp\left[-g^2 \hbar t^2 \frac{M \hbar}{2\left(\frac{M}{N}+ \frac{N}{M}\right)} \right]\nonumber\\
  &\times&\cos\left(g \hbar\sqrt{N^2+M^2}t \right).
\end{eqnarray}

We see that the argument of the cosine $ g \hbar\sqrt{N^2+M^2} t = g
J_{\rm tot} t$ involves only the classical value of the total angular
momentum $J_{\rm tot}$, while the decay exponent involves two powers
of $\hbar$, of which only one is absorbed into the definition of
classical angular momenta. The decay exponent is a quantum
effect, in a sense that its physical origin lies in the quantum
uncertainty principle.
This allows us to make an important identification: coherent
evolution in the neutrino system maps into the classical behavior of
the angular momenta, while incoherent effects in the neutrino system
correspond to quantum effects in the spin system. This important
point will be developed further in Sect.~\ref{sec:freeze-out}.

Finally, we note that the solution is periodic and the wavetrain,
having seemingly completely decayed away, reemerges after some time.
This phenomenon is due to the fact that the spin system possesses a
fundamental frequency, of which all other frequencies (the arguments
of the cosines in Eq.~(\ref{eq:form of probability})) are multiples.
This effect will be discussed in detail in Sect.~\ref{sec:period}.

\section{Analysis: A flavor eigenstate background}
 \label{sec:eigenstate_bg}

\begin{figure*}[htbp]
\psfrag{t}{$\tau$}
\begin{center}
   \fbox{\epsfig{file=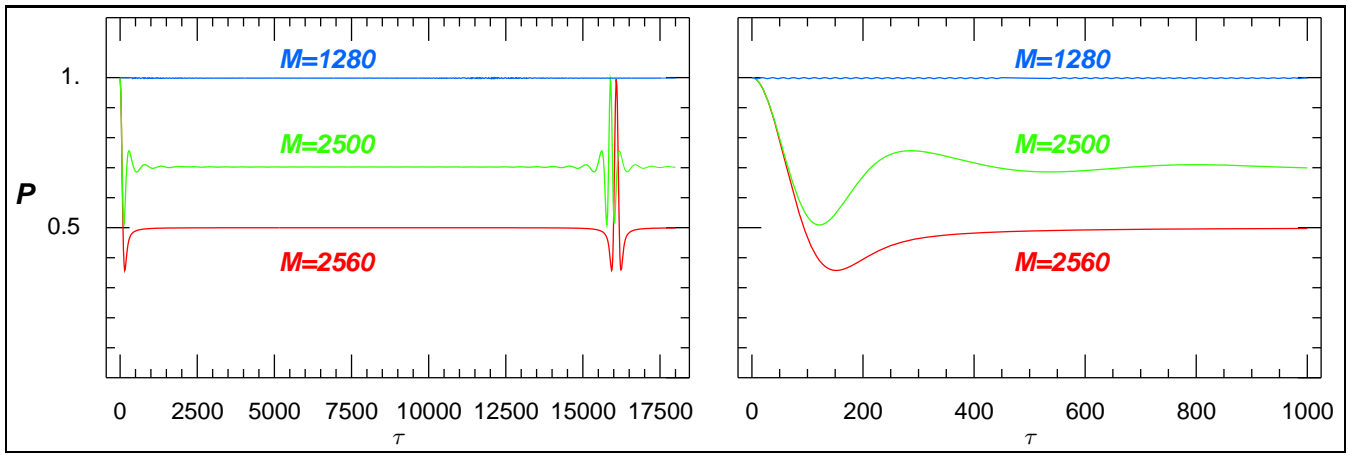, height=5.7cm, angle=0}}
\end{center}
\caption{Plots of N=2560 and various numbers of spin down
particles, M.  The time variable $\tau$
is defined as the scaled time $\tau = gt(M+N)$. Note that for all the
graphs $N+M= $ even.  The right figure shows that the equilibration
time for the bottom two curves is characteristic of incoherent
evolution, $t_{eq} \propto [2/(M+N)]^{1/2}$.}
\label{fig:N2560 Plot}
\end{figure*}

We next analyze the probability for the case that the state $j_M$ is
made of spin down particles (or neutrinos in the muon flavor
eigenstate). Hence we identify the quantum numbers as
\begin{equation}
\label{eq:quantum numbers1}
\begin{split}
m_N=\frac{N}{2}, \hskip0.3truecm  m_M =-\frac{M}{2}.
\end{split}
\end{equation}
One reason this case is interesting for us is that for this system
the one-particle formalism predicts no coherent flavor conversion.
Indeed, the off-diagonal terms in the Hamiltonian
(\ref{eq:H_oneparticle}) vanish.

The explicit form of the coefficients $C(J)$ and $\eta(J)$ is given in
Appendix \ref{sec:app_c_eta}, Eqs.~(\ref{eq:C2_eigenstates}),
(\ref{eq:C2_eigenstates_zero}), and (\ref{eq:eta2_eigenstates}).  As
explained there, for the case $N=M$ our probability agrees exactly
with that found in F\&L II.

We have plotted the probability according to these equations for
various numbers of spin down ($M$) and spin up ($N$) particles and a
subset of these is shown in Fig.~\ref{fig:N2560 Plot}. The time on
the figures is scaled such that $\tau = tg(N+M)$ so that we may
compare our results to those in F\&L II and BRS.

\begin{figure}[h]
\begin{center}
   \fbox{\epsfig{file=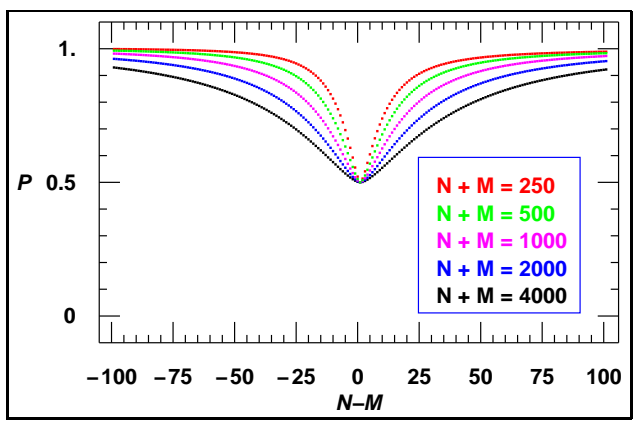, height=5.2cm, angle=0}}
\end{center}
\caption{The average probability ($\bar{P}_1=1/2+\sum_J C(J)$) when we change
the number of spin ups and spin downs while keeping $M+N$ constant.}
\label{fig:average probability}
\end{figure}

\begin{figure}
\psfrag{t}{$\tau$}
\begin{center}
   \fbox{\epsfig{file=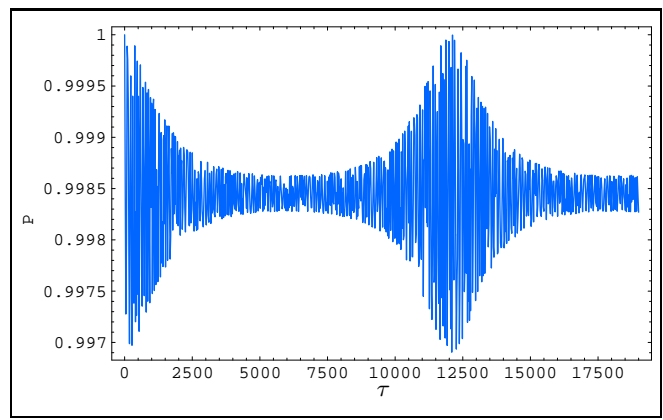, height=5.2cm, angle=0}}
\end{center}
\caption{A close up of the plot for N=2560 and M=1280.  Note the
vertical scale.  The probability is very close to one and fluctuates
minimally about the average value. } \label{fig:Plot M1280}
\end{figure}

The main features of the solution we find are the following:
\begin{itemize}
\item As seen in the left panel of Fig.~\ref{fig:N2560 Plot}, for
  $N\sim M$ (to be quantified later), the system equilibrates to some
  value of $\bar{P}_1$ and subsequently for a long time shows no
  evolution. This behavior is similar to what was observed in the case
  of Sect.~(\ref{sec:superposition_bg}). Just like that system, the
  wavetrain reemerges after some time.
\item As seen in the right panel of Fig.~\ref{fig:N2560 Plot}, the
  equilibration happens on time scales characteristic of incoherent
  evolution, $t_{eq} \sim g^{-1}((M+N)/2)^{-1/2}$. Unlike the case of
  Sect.~(\ref{sec:superposition_bg}) though, in the present case for
  $N=M$ we do not see coherent oscillations with the decaying
  wavetrain.  In fact, it was explicitly shown in F\&L II that for
  $N=M$ the probability $P_1$ depends on time only through the
  combination $\sqrt{N}g t$. Thus, the evolution has manifestly
  incoherent nature.  ({\it cf} Eq.~(\ref{eq:coh_analyt})).
\item As $|N-M|$ is increased, the value of $P_1$ in equilibrium,
  $\bar{P}_1$, increases and for large $|N-M|$ the system stops
  evolving, entering a ``freeze-out" state. This trend can be clearly
  seen in Fig.~\ref{fig:N2560 Plot}. As further illustrated in
  Fig.~\ref{fig:average probability}, where $\bar{P}_1$ is plotted as
  a function of $N-M$ for fixed $M+N$, the minimum value of
  $\bar{P}_1=1/2$ occurs when $N=M$. That $\bar{P}_1$ rises \emph{for
    both signs} of $(N-M)$ appears counterintuitive at first. Indeed,
  it means if we have a single spin up coupled to a very large system
  of spins down, the first spin does not equilibrate to a state mostly
  oriented down, but remains aligned up.
\item As $|N-M|$ is increased, in addition to decay, the system start
  exhibiting oscillations, the frequency of which grows with $|N-M|$.
  The amplitude of these oscillations gets progressively smaller as
  $|M-N|$ increases, since the system approaches the freeze-out state.
  This behavior can be seen in Fig.~\ref{fig:Plot M1280}, which
  reproduces at higher resolution the case $M=1280$, $N=2600$ of
  Fig.~\ref{fig:N2560 Plot}.
\end{itemize}

In what follows, we will discuss these features further. This will
allow us to gain a deeper understanding of the spin system and its
relationship to the neutrino system. We will first show how the
transition from the regime in which the spins equilibrate to the
freeze-out regime illustrates the interplay of ``classical" and
``quantum" effects. We will then discuss the time scale of the
evolution and whether the system follows the predictions of the
one-particle Hamiltonian in the classical regime. Finally, in
Sect.~\ref{sec:period} we will comment on the periodicity of our
solution.

\subsection{From equilibration to freeze-out: a quantum-to-classical transition}
\label{sec:freeze-out}

In the course of the analysis in Sect.~\ref{sec:superposition_bg}, we
have encountered a situation in which fast coherent processes in the
neutrino system corresponded to classical effects in the spin system,
while slower incoherent processes in the neutrino system corresponded
to quantum effects in the spin system. This physical picture is
further illustrated by the system $m_N=N/2$, $m_M=-M/2$ we are now
considering.

Once again, we recall that this system is reduced to two angular
momenta as described in Sect.~\ref{sect:methodII}. When $N$ and $M$
are large, these two angular momenta become approximately classical
objects, $\vec{j}_1$ and $\vec{j}_2$. These objects are coupled with
an interaction $const \times \vec{j}_1 \cdot \vec{j}_2$. By energy
conservation, this quantity stays fixed, and since the lengths of
both vectors are fixed the angle between the vectors $\vec{j}_1$ and
$\vec{j}_2$ stays fixed. Since the two vectors start out pointing in
opposite directions, they remain pointing in opposite directions.
The only possibility left is for the system to tilt as a whole, but
that would violate momentum conservation, \emph{unless} the total
momentum of the system is \emph{zero} ($N=M$). Thus, the classical
system is frozen, unless $j_1=j_2$, which is a special point.

Our calculation shows that quantum mechanical effects actually
resolve this discontinuity. The transition from the freeze-out to
equilibration happens in some finite range of angular momenta.

What physics sets this range? Let us consider the two angular momenta
being added as semiclassical objects. Then, in addition to the angular
momenta in the $z$ direction, they each possess ``quantum'' angular
momenta in the $x-y$ plane. These momenta are of the order
$$J_{xy}\sim\sqrt{J^2-J_z^2}=\sqrt{J (J+1)-J^2}=\sqrt{J}.$$
That this is a quantum effect is seen by restoring the units:
$J_{xy}\sim \sqrt{J\hbar}$. This ``quantum'' angular momentum is
what is responsible for the equilibration.

When the two angular momenta are added, their $x-y$ quantum momenta
are also combined and the net object has $J_{x-y}$ of the same order
as the ingredients. If the net classical momentum along the $z$ axis
is greater then $J_{x-y}$, the equilibration will be suppressed. In
other words, we arrive at the physical condition that determines the
boundary between equilibration and ``freeze-out''
\begin{equation}
  \label{eq:phys_equil}
  |M-N|/2 \sim \sqrt{(M+N)/2}.
\end{equation}
For large number of spins, we can simply replace in this condition
$\sqrt{(M+N)/2} \rightarrow \sqrt{N}$.

This boundary between freeze-out and equilibration can be quantified
by the analytical expression for the equilibrium $P_1$ that can be
derived for $|M-N|\ll (M+N)$ (see Appendix \ref{sect:appB}):
\begin{eqnarray}
  \label{eq:analyt_equil}
  \bar{P}_1 \simeq 1 - \frac{1}{1+\Delta/\Sigma}
  F\left(\frac{\Delta}{\sqrt{\Sigma}} \right),
\end{eqnarray}
Here, $\Delta\equiv (N-M)/2$, $\Sigma\equiv (M+N)/2$ and
\begin{eqnarray}
   F(a) \equiv \frac{1}{2}-\frac{a^2}{2} \int_{0}^{\infty} dy
    \frac{e^{-  y}}{y+a^2}.
\end{eqnarray}

Up to a small correction in the prefactor in
Eq.~(\ref{eq:analyt_equil}), $\bar{P}_1$ depends on $\Delta$ and
$\Sigma$ in combination $\Delta^2/\Sigma$. Moreover, $F(a)$ quickly
drops to zero as $a$ is increased beyond one. Hence, the width $|M-N|$
of the ``quantum equilibration region'' indeed scales with $N+M$ as
$\sqrt{N+M}$.

This scaling can be seen in Fig.~\ref{fig:average probability}. The
curves shown there are obtained by using the exact solution,
Eqs.~(\ref{eq:form of probability}), (\ref{eq:C2_eigenstates}),
(\ref{eq:eta2_eigenstates}), (\ref{eq:C2_eigenstates_zero}). They are
well described by Eq.~(\ref{eq:analyt_equil}).

\subsection{Equilibration Time}
\label{sect:subtlety}

We now discuss the time scales that control the evolution of our
system. We are particularly interested to see how the evolution scales
with the number of spins in the limit when the spin system is large.

As already mentioned, the case of $N=M$ considered in F\&L II is
straightforward: for large $N$ the evolution is uniquely dependent on
$\sqrt{N}t$ and the decay curve is
\begin{eqnarray}\label{eq:FLIIintegral}
    P_1(t)=1-\frac{\sqrt{\pi}}{2}\sqrt{N}g t\exp(-N g^2 t^2)
    {\rm erfi}(\sqrt{N}g t),
\end{eqnarray}
where ${\rm erfi}(z)$ is the imaginary error function
\begin{equation}
{\rm erfi}(z)\equiv -i~{\rm erf}(iz)=
\frac{2}{\sqrt{\pi}}\times\int_0^z \exp(t^2)dt~. \label{erfi}
\end{equation}

Our investigations in Sect.~\ref{sec:freeze-out}, however, showed
that in general the situation is less obvious. For $N\ne M$
Figs.~\ref{fig:N2560 Plot}, \ref{fig:Plot M1280} show a
more complicated evolution pattern. Indeed, as we show next, in
this case a new, shorter time scale enters the evolution.

First, let us consider the simplest example, an obvious limiting case
$N=1$. As can be easily seen from Eq.~(\ref{eq:form of probability}),
the evolution contains just a single oscillation frequency, as $J_{\rm
  min}=M/2-1/2$ and $J_{\rm max}=M/2+1/2$. The corresponding value of
$\eta(M/2-1/2)$ is $2/(M+1)^2$. We thus have
\begin{equation}\label{eq:single_spin_eigenstate}
    P_1(t)= 1-\frac{2}{(1+M)^2}+\frac{2}{(1+M)^2} \cos [g t (M+1)].
\end{equation}
We see that the oscillation timescale grows linearly with the number
of spins, not as a square root as in Eq.~(\ref{eq:FLIIintegral}). We
also notice that the amplitude of the oscillation is suppressed by
the second power of the number of spins.

\begin{figure}[htbp]
  \centering
  \includegraphics[width=0.47\textwidth]{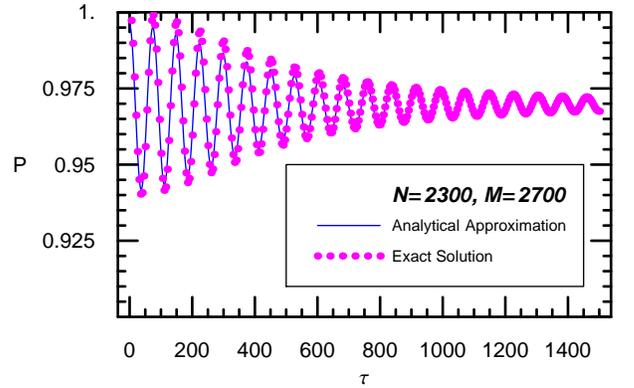}
  \caption{Time evolution of $P_1(t)$ in the case  $N=2300$, $M=2700$,
  $m_N=N/2$, $m_M=-M/2$. The Figure illustrates that
  Eq.~(\ref{eq:asympt_firstterm}) (curve) provides a very good approximation
  to the exact solution obtained by summing
  the series in Eqs.~(\ref{eq:form of probability}),
  (\ref{eq:C2_eigenstates}), (\ref{eq:eta2_eigenstates}) (dots).
  The time variable is defined as $\tau\equiv g t (M+N)$, the same way
  as in Figs.~\ref{fig:N2560 Plot}, \ref{fig:Plot M1280}.}
  \label{fig:asymptotic}
\end{figure}

Next, let us consider a more nontrivial case
$\sqrt{N+M}<|N-M|\ll(N+M)$. This case is characterized by the decay of
the wavetrain to the value $\bar{P}_1$ given in
Eq.~(\ref{eq:analyt_equil}). It turns out that for
$\sqrt{N+M}<|N-M|\ll(N+M)$ the decay behavior is qualitatively
different from Eq.~(\ref{eq:FLIIintegral}). The wavetrain in this case
is approximately described by
\begin{eqnarray}
    \label{eq:asympt_firstterm}
  P_1(t) &\simeq& 1-\frac{\Sigma}{2}\left(\frac{1}{\Delta^2} -
  \frac{\cos\left(2 g t \Delta + 2\arctan\left(\frac{gt\Sigma}{\Delta}\right)\right)}{\Delta^2+g^2t^2\Sigma^2}
  \right).
  \;\;\;\;\;
\end{eqnarray}
For details see Appendix~\ref{sect:app_asympt_series}, where  an
asymptotic series for $P_1(t)$ is derived.
Eq.~(\ref{eq:asympt_firstterm}) contains the leading term of this
series. For sufficiently large $|M-N|$ (compared to $\sqrt{N+M}$) it
gives an excellent approximation of the true result, as can be seen
in Fig.~\ref{fig:asymptotic}.

By inspecting Eq.~(\ref{eq:asympt_firstterm}), we clearly see that
for nonzero $|M-N|$ a new frequency, $2 g \Delta = g (N-M)$ enters
the problem. We stress that this frequency depends on $N-M$
linearly. The amplitude of the oscillations at the maximum is
suppressed by $(N+M)/|N-M|^2$, so as a function of $|M-N|$ the
oscillations increase in frequency and decrease in amplitude.
Finally, the oscillations decay with time as only a power law and
the decay time, $t_{eq}\sim |N-M|/(g(N+M))$ increases with $|M-N|$.
Compare this with the situation for $N=M$, in which case the decay
time scale is $t_{eq}\sim 1/(g\sqrt{N})$ and the decay has
exponential dependence.

At last, what conclusions can be drawn about the applicability of
the one-particle coherent Hamiltonian in this case? Our conclusions
is that one can define in what sense it is applicable, but the
argument is a bit subtle. Unlike in the case $N=M$ considered in
F\&L II, in the more general setup considered here we find a new
frequency, linearly dependent on the difference $|N-M|$. Can this be
reconciled with Eq.~(\ref{eq:H_oneparticle}), which states that in
the neutrino system which begins as a collection of flavor
eigenstates there should not be any flavor coherent conversion? We
recall that while the frequency of our solution increases with
$|N-M|$, the amplitude of the corresponding oscillations becomes
smaller and smaller (see Eq.~(\ref{eq:asympt_firstterm})). The
oscillations become high-frequency as the system enters a state of
freeze-out. The one-particle coherent limit corresponding to
neglecting the residual small oscillations.

A different way to state this is that the oscillation amplitude is
suppressed by $(M+N)/(N-M)^2$ and becomes large only when $|N-M|$
approaches $\sqrt{N+M}$. In the latter case, however, the
oscillation frequency, obviously, becomes of order $\sqrt{N+M}$, and
hence indistinguishable from the incoherent time scale.

\subsection{The Period of the Probability}
 \label{sec:period}
In this section we study the periodicity of the probability $P_1(t)$
observed in Fig.~\ref{fig:N2560 Plot}.
Recall, the probability is
\begin{equation}
\label{eq:period eqn}
P(t)=const + \sum_{J=|j_N-j_M|}^{j_N+j2-1} \eta(J)\cos\left[gt\thickspace(2J+2)\right].
\end{equation}
The periodicity is a consequence of the probability being a sum
of cosines with frequencies that are multiples of the lowest,
fundamental frequency.

\subsubsection{The Case for N=1 or M=1}

We first consider the case when N or M is one. Consider $N=1$. For
this case there is only one cosine term with $J=\frac{M}{2}-\half$.
Therefore Eq. (\ref{eq:period eqn}) reduces to,

\begin{equation}
P(t)=\sum_{J=|j_N-j_M|}^{j_N+j2} C(J) + \eta\left(\frac{M-1}{2}\right)\cos\left[gt\thickspace(M+1)\right]
\end{equation}

Hence,

\begin{equation}
T_{N=1}=\frac{2\pi}{g(M+1)}
\end{equation}

If we take $M=1$ instead, the result is the same except N and M are swapped.

\subsubsection{The Case for $N>1$ and $M>1$}

 Each cosine in Eq.(\ref{eq:period eqn}) satisfies,

\begin{equation}
\cos\left[gt\thickspace(2J+2)\right] = \cos\left[g(t + T_J)(2J+2)\right]
\end{equation}

Here $T_J$ is the period of the cosine corresponding to angular
momenta $J$.  Now for all $J$,

\begin{equation}
\label{eq:cos period}
T_J = \frac{2\pi}{g(2J+2)}
\end{equation}

To find the period we need the least common multiple of the $T_J$'s.
We find the period (together with the previous subsection) to be,

\begin{equation}
\label{eq:period}
T=
\begin{cases}
\frac{\pi}{g}, &\text{ if $N+M=$ even (or J=integer)}\\
\frac{2\pi}{g}, &\text{ if $N+M=$ odd (or J= half integer)}\\
\frac{2\pi}{g(N+M)}, &\text{ if N=1 or M=1}
\end{cases}
\end{equation}

Note that the period is the same for the case of a flavor eigenstate
background and a flavor superposition background because the period
only depends on the argument of the cosine.  The argument of the
cosine is the same for both cases. The discontinuity between the
period when $N=1$ or $M=1$ and the other cases arises because in the
first two cases there is an interference of many cosine waves (as
there the cosines are summed over) and in the last case there is
only one cosine wave. For $M=N$, our results reduce to that found in
F\&L II.

Notice that, up to the factor of two controlled by whether $N+M$ is
even or odd, the period of the spin system depends only on the value
of the spin-spin coupling $g$ and not on the size of the spin
system. For a neutrino system, $g=\frac{\sqrt{2} G_F}{V}$ and the
period seems to depend on the volume occupied by the neutrino
system. In fact, in a real neutrino gas, non-forward scattering
effects would destroy any periodicity. We recall that for
sufficiently large numbers of neutrinos the time scale of incoherent
forward scattering, Eq.~(\ref{eq:t_inc}), is much smaller than the
period found here and, moreover, that forward incoherent scattering
is only a small fraction of the total incoherent scattering (see
F\&L II). Thus, the periodicity is an example of an effect in the
spin system that is not realized in the actual neutrino system. In
the context of the spin system, on the other hand, the effect is
perfectly physical and indeed is reminiscent of the well-known
``spin echo" effect experimentally observed in actual spin systems.

\begin{figure}[b]
\begin{center}
   \fbox{\epsfig{file=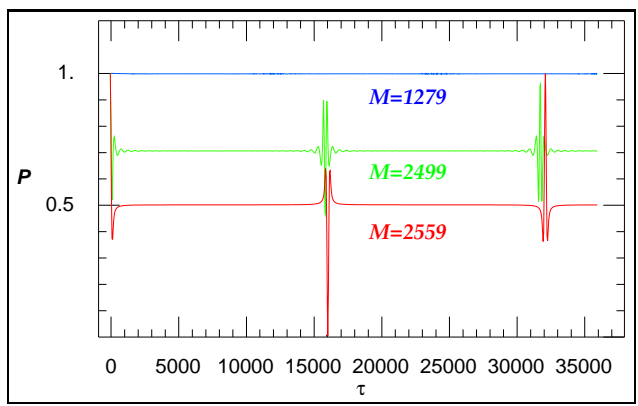, height=5.0cm, angle=0}}
\end{center}
\caption{Plots of $N=2560$ and various numbers of spin down
particles, $M$.  The time is scaled so that $\tau = gt(M+N)$. Note
that for all the graphs $N+M=$odd} \label{fig:NModd Plot}
\end{figure}

As a final comment note that when we approximate the sums over the
cosines by the integrals, in Eqs.~(\ref{eq:coh_analyt}),
(\ref{eq:analyt_equil}), and (\ref{eq:asympt_firstterm}), the
periodicity of the solution is destroyed. This has to do
with the disappearance of the fundamental (lowest) frequency in the
system. The same observation was made in F\&L II for $N=M$.

\subsubsection{ A note about minima}

Let us call the case when $P_1=1/2+\sum_J C(J) -\sum_J \eta(J)$
(which is the lowest possible value that the probability can be) a
perfect minimum. For the perfect minimum to occur we must have,
$\cos\left[gt \thickspace(2J +2)\right] = -1$ simultaneously for all
$J$. The period of each cosine decreases as $J$ increases, therefore
we only need to find the time when the cosine with the largest
period and the cosine with the smallest period are simultaneously
equal to $-1$.

By a calculation analogous to that for the period  one can show that
if $N+M=$ even there will never be a perfect minimum.  If $N+M$ is
odd the times when the probability attains a perfect minimum is,

\begin{equation}
T_{min} = \frac{k\pi}{g}
\hskip0.5truecm {\rm where} \hskip0.5truecm k= 1,2,3 ...
\end{equation}

Note that this is half way between perfect maxima (the case where
$P_1(t)=1/2+\sum_J C(J) + \sum_J \eta(J)$). Hence if $M+N$ is odd
(or $J$ is half an integer) the probability has  both a perfect
minimum and a perfect maximum. This result shows yet another
intriguing physical feature of the spin system. While a system with
$N+M=$ even is characterized by a set of recurring maxima, in the
system with $N+M=$ odd every other such maximum is replaced by a
minimum.
This behavior is illustrated in Fig.~\ref{fig:NModd Plot}. Note that
in this figure, as before, the abscissa is the scaled time $\tau =
gt(M+N)$.


\section{The freeze-out effect and a real neutrino gas: a critical analysis}
\label{sect:frezeout_musings}

\subsection{Overview of the problem}

In Sect.~\ref{sec:eigenstate_bg} we found that a general
``$N+M$'' ($N\ne M$) spin system that starts polarized in the $z$
direction typically equilibrate very little, even on the longer
incoherent time scale, except in cases when $N$ and $M$ are very
close ($|N-M|\lesssim \sqrt{N}$). Obviously, it is important to
understand if this finding corresponds to the behavior of a real
neutrino gas. This question is not a trivial one. As mentioned in
the introduction, going from a neutrino gas to the spin model
involves several simplifying assumptions: only forward scattering is
kept, equal interaction strengths are assumed, the interactions are
taken to be continuous in time, etc. All these assumptions could, in
principle, introduce certain artifacts in the system and it is not
{\it a priori} obvious that the freeze-out behavior found in the toy
model is not just such an artifact. We will examine this question
next.

\subsection{Interference of neutrino scattering amplitudes}
\label{sect:critical_idea}

Consider a test neutrino, taken for definiteness to be $|\nu_e>$,
flying though a box with the gas of neutrinos. Assume the neutrinos
in the gas are in the flavor eigenstates,
\begin{equation}\label{eq:initstate_nu}
    |S(0)\rangle=|\underbrace{\nu_e \nu_e...\nu_e}_{(N-1)}
    \underbrace{\nu_\mu\nu_\mu...\nu_\mu}_M\rangle~.
\end{equation}
Let us limit scattering to only the forward process and ask: what is
the probability that scattering changes flavor of the test neutrino?

Let us review the arguments of F\&L I. Omitting the
flavor-preserving neutral current piece, which is the same for all
neutrinos, we can write the result of the interaction to first order
in the interaction amplitude $a$ as
 \begin{widetext}
\begin{eqnarray}\label{eq:exchange}
    \lefteqn{|\nu_e\rangle |\underbrace{\nu_e \nu_e...\nu_e}_{(N-1)}
    \underbrace{\nu_\mu\nu_\mu...\nu_\mu}_M\rangle \longrightarrow} \nonumber\\
    |F\rangle &=&(1+(N-1)ia)|\nu_e\rangle |\nu_e \nu_e...\nu_e
    \nu_\mu\nu_\mu...\nu_\mu\rangle\nonumber\\
    &+&
    i a(\underbrace{|\nu_\mu\rangle |\nu_e \nu_e...\nu_e
    \nu_e\nu_\mu...\nu_\mu\rangle+
    |\nu_\mu\rangle |\nu_e \nu_e...\nu_e
    \nu_\mu\nu_e...\nu_\mu\rangle+ ...}_M),
\end{eqnarray}
\end{widetext}
The interaction amplitude $a$ is given by $a\equiv C\delta t\equiv
-\sqrt{2}G_F(1-\cos\Theta)\delta t/V$
\cite{Friedland_&_Lunardini_1,Friedland_&_Lunardini_2}, where
$\delta t$ is the duration of the interaction between a pair of
neutrinos. Since the final states are mutually orthogonal, the
probability of flavor change goes as $M a^2$ indicating the
incoherent nature of the evolution.

Eq.~(\ref{eq:exchange}) assumes that the time step $\delta t$ is
small such that not only $C\delta t\ll 1$, but also $N C\delta t\ll
1$ are satisfied. While the former is indeed true for any reasonable
neutrino system of interest \cite{Friedland_&_Lunardini_2}, the
latter need not hold, as $N a$ or $Ma$ for large $N$ and $M$ are not
necessarily small. To make contact with the spin system, which stays
continuously coupled, we must consider longer interaction times.
Correspondingly, we need to modify Eq.~(\ref{eq:exchange}),
\emph{keeping higher powers of $a$}.

One can accomplish this by regarding (\ref{eq:exchange}) as the
first order term in the perturbative expansion of the wavefunction
at the time $\delta t$ ($N C\delta t\ll 1$, $M C\delta t\ll 1$), and
using this to make further steps in time to the time $T=n\delta t$
($C T\ll 1$, but not necessarily $N C T\ll 1$, $M C T\ll 1$). Let us
define a convenient notation of the initial state
\begin{equation}
|I\rangle =  |\nu_e\rangle |\underbrace{\nu_e \nu_e...\nu_e}_{(N-1)}
    \underbrace{\nu_\mu\nu_\mu...\nu_\mu}_M\rangle
\end{equation}
and the ``exchange" state
\begin{eqnarray}
|E\rangle &=& \underbrace{|E_1\rangle + |E_2\rangle + ...}_M,\\
|E_1\rangle &=& |\nu_\mu\rangle |\nu_e \nu_e...\nu_e
    \nu_e\nu_\mu...\nu_\mu\rangle,\\
|E_2\rangle  &=& |\nu_\mu\rangle |\nu_e \nu_e...\nu_e
    \nu_\mu\nu_e...\nu_\mu\rangle,\\
    ...&&\nonumber
    \end{eqnarray}

Then at the time $T$ the state is
\begin{widetext}
\begin{eqnarray}
|T= n\delta t\rangle &= & \left(1 + i(N-1)C\delta t\right)^n|I\rangle  \nonumber
\\
& +& iC \delta t \left[\sum_{r=0}^{n-1} \left(1 + i(N-1)C\delta t\right)^{n - 1 - r}\left(1 + i(M-1)C\delta t\right)^r\right] |E\rangle \nonumber \\
&+ & O([iC\delta t]^2),
\end{eqnarray}
\end{widetext}
as can be readily verified by induction.

The cyclotomic polynomial is  summed to give
\begin{widetext}
\begin{eqnarray}
|T = n\delta t\rangle &= &  \left(1 + i(N-1)C\delta t\right)^n|I\rangle \nonumber \\
& +& iC\delta t \frac{\left(1 + i(N-1)C\delta t\right)^n - \left(1 +
i(M-1)C\delta t\right)^n}{iC(N-M)\delta t} |E\rangle
\nonumber \\
& +&  O([iC\delta t]^2),
\end{eqnarray}
\end{widetext}
and, taking the limit $n \to \infty$
\begin{eqnarray} \label{eq:sumcyclotomic}
\nonumber |T=n\delta t\rangle &=& e^{i(N-1)CT} |I\rangle \\\nonumber
&+& \frac{\left(e^{i(N-1)CT} - e^{i(M-1)CT}\right)}{(N-M)}
|E\rangle\\ &+&O([iC\delta t]^2).
\end{eqnarray}
Note that the explicit dependence on $\delta t$ has now disappeared.
This is to be expected since the formal solution of the
Schr\"odinger equation for the system depends only on the time
difference between the initial and final times.

The reason for the freeze-out behavior  is now clear.  For large
values on $N-M$, the probability of a transition to the exchange
state is suppressed by the factor $(N-M)^{-2}$.

Notice first of all that this effect disappears when $N=M$. This is
why in this case one can observe full equilibration and test the
incoherent nature of the evolution.  However, one should note that
in the present approximation the $N \to M$ limit gives $iCT$  as the
amplitude for the state $|E\rangle$.  As in this case the phase of
the initial and exchange states are the same, we regard one factor
of $T$ in the probability as a relic of the energy conservation
delta function, and obtain as the probability per unit time for a
transition from the initial state
\begin{equation}
\frac{d P}{d t} = M |C|^2
\end{equation}
(remember that $\langle E|E \rangle = M$), showing incoherent equilibration.

\subsection{Does the freeze-out happen in a real neutrino system?}

Finally, we can address the key question: does the freeze-out
observed in the model of interacting spins actually happen in a real
neutrino system? The answer is negative.

Let us return to the arguments in the previous subsection. Upon
closer inspection of Eq.~(\ref{eq:exchange}), it becomes clear that
for large $N$, $M$ and times $T\ll C ^{-1}$, the problem is in fact
equivalent to an oscillation set-up with a Hamiltonian that, in the
basis $|I\rangle, |E_1\rangle, |E_2\rangle, ...$, has $(i C N, i C
M,i C M,i C M,...)$ on the diagonal\footnote{We assume large $N$ and
$M$ and neglect terms of order one.} and factors of $i C$ on the
off-diagonal, in the first column (and the corresponding terms in
the first row). For the case $N=M$ considered in
\cite{Friedland_&_Lunardini_2}, the transition probability between
$|I\rangle$ and any of the states $|E_i\rangle$ is given by $P\sim
|i t  C|^2$. The generalization of this to $N\ne M$ is
\begin{eqnarray} \label{eq:phaseintegral}
   P\sim \left\arrowvert i\times C \times \int_0^T dt \exp[i (N-M) t \times C]\right\arrowvert^2.
  \end{eqnarray}

It is straightforward to check that the transition probability that
follows from Eq.~(\ref{eq:phaseintegral})
 \begin{eqnarray}
\nonumber P &\sim&
\left\arrowvert\frac{\exp[i C (N-M) T ]-1}{ (N-M)}\right\arrowvert^2\\
&=& \frac{2-2\cos((N-M)CT)}{(N-M)^2}
\end{eqnarray}
agrees with that obtained from Eq.~(\ref{eq:sumcyclotomic}),
indicating that the corresponding amplitudes are the same, up to an
overall phase convention.

Let us take the time $T$ in Eq.~(\ref{eq:phaseintegral}) to be the
timescale of incoherent equilibration, given by Eq.~(\ref{eq:t_inc}):
\begin{equation}\label{eq:tincFL2}
    t_{\rm eq}^{inc} \sim (G_F \sqrt{(N+M)/2}/V)^{-1}.
\end{equation}
If $N=M$, the probability to transition into any one of the exchange
states $|E_i\rangle$ in this time is $P\sim (C t_{\rm eq}^{inc})^2$.
Remembering that $C=-\sqrt{2}G_F/V(1-\cos\Theta)$ and dropping the
angular factor $(1-\cos\Theta)$ (which is also dropped in the
derivation of Eq.~(\ref{eq:tincFL2})), we get $P\sim((N+M)/2)^{-1}$.
The probability of transition into any of the exchange states is
$P(|I\rangle\rightarrow|E_{any}\rangle)\sim (N-1+M) P \sim 1/2$, thus
confirming that the equilibration does happen on the time scale
$t_{\rm eq}^{inc}$.

If, however, we have $(N-M)CT \gtrsim 1$, the oscillating exponent in
Eq.~(\ref{eq:phaseintegral}) suppresses the transition and the system
freezes out. Thus, we immediately obtain a physical estimate for the
freeze-out:
\begin{equation}\label{eq:freeze_cond2}
    |N-M| \sqrt{2/(N+M)} \gtrsim 1,
\end{equation}
which is nothing but the condition we found in the spin system,
Eq.~(\ref{eq:phys_equil}), up to a trivial factor. This provides a
powerful check of the validity of Eq.~(\ref{eq:phaseintegral}) and the
physical picture that led to it.

We can now give interpret Eq.~(\ref{eq:phaseintegral}) the following
interpretation.  For $N$, $M\gg 1$, the states $|I\rangle$,
$|E_i\rangle$ are \emph{approximately} energy eigenstates. Strictly
speaking, of course, the true energy eigenstates are mixtures of
$|I\rangle$, $|E_i\rangle$, but for large $N$, $M$ one of the
eigenstates is predominately $|I\rangle$, with a small admixture of
the other states, and vice versa. Correspondingly, we can loosely
speak of the energies of the $|I\rangle$ and $|E_i\rangle$ states. For
$N\ne M$ the state $|I\rangle$ and any of the states $|E_i\rangle$
have different energies.  The incoherent spin flips in the spin system
are thus forbidden by energy conservation (enforced by the oscillating
exponent in Eq.~(\ref{eq:phaseintegral})).

In a real neutrino system, of course, there is an additional degree
of freedom: the neutrino momentum. The energy liberated in the
process of flavor exchange is converted into a slight shift of the
neutrino kinetic energy and the energy conservation is assured. In
the simplest manifestation of this point, a neutrino existing a
dense neutrino region will gain a small amount of kinetic energy.
(The neutrino-neutrino interactions, just like any interactions
between like particles mediated by a vector boson, are repulsive.)

Thus, the freeze-out observed in the spin model is an artifact of
the model, more specifically, a consequence of the dropped degrees
of freedom, neutrino momenta.

This discussion applies to the system of neutrino plane waves in a
box, which was the system for which the spin model was created. It
is instructive to instead consider a system of interacting neutrino
wavepackets. In this system the interaction energy is localized to
the regions occupied by the overlapping wavepackets, with the total
interaction energy (averaged over volume) being the same as in the
case of plane waves. The statement of the changing kinetic energy,
as the neutrino moves from a more dense to a less dense region, has
its analogue in the case of wavepackets as well: The kinetic energy
of a given packet changes as it interacts with another packet. Upon
spatial averaging this change, one gets the ``plane waves" result.
There are fewer interactions in less dense regions, hence the
averaged kinetic energy there is higher.

The advantage of thinking about wave packets is that in this case it
is clear that the rate of incoherent scattering cannot be
suppressed: for sufficiently small wavepackets they interact with
each other two-at-a-time, giving the usual scattering rate. The
answer in the physical system cannot depend on the size of the
wavepackets, since the interaction Hamiltonian is
energy-independent.

\section{Conclusions}

No problem in physics can be solved exactly. Often, however, a given
physical problem could be mapped onto an idealized system, for which
a complete solution exists. One then needs to critically examine
which features of the solution carry through to the original
physical system and which arise as artifacts of the simplifying
assumptions made along the way.

We have performed just such an analysis. Our goal was to investigate
coherent effects in a gas of neutrinos. We have simplified the system
by limiting the scattering to only forward direction and omitting the
momentum degrees of freedom. With only the flavor degrees of freedom
remaining, the neutrino system could be mapped onto a system of
coupled spins. We further simplified the problem by setting the
coupling strengths to be equal, and considering a certain class of
initial states. We have shown that the resulting spin model could be
solved \emph{exactly}. Our solution generalizes the analysis of F\&L
II \cite{Friedland_&_Lunardini_1}, where a particular initial state,
the one with equal numbers of up and down spins, was considered.

The solution proved to be very instructive. The spin system
exhibited both coherent oscillations and incoherent decay,
reproducing effects expected for a real neutrino gas. Several
examples were considered in Sects.~\ref{sec:superposition_bg} and
\ref{sec:eigenstate_bg}. As these examples demonstrate, coherent
effects in the neutrino system correspond to the behavior of the
spin system in the classical limit; likewise, incoherent effects in
the former have their analogue in quantum effects in the latter.

We have presented two different methods of obtaining the solution.
The first construction separates the system in three parts: the test
spin, the other spins that started in the same orientation and the
remaining spins. The second method exploits the high degree of
symmetry of our chosen initial states. Both methods have complementary
advantages. The second of the methods provided crucial physical
intuition for understanding the classical limit of the system and the
interplay of quantum and classical effects, as shown in
Sects.~\ref{sec:superposition_bg} and \ref{sec:freeze-out}. At
the same time, the first method is more general and can be used to
generate solutions for any initial state of the test spin (by simply
changing the state of the test spin by adjusting the parameter
$\lambda$ in Sect.~\ref{sec:split_spin_P1} ).

Several interesting results were found. One such result involves the
subtlety of identifying coherent effects in the neutrino system by
taking a classical limit of the spin system. The special case
considered in \cite{Friedland_&_Lunardini_1} seemed to suggest that in
the large $N$ quantum effects separate from classical ones necessarily
by having longer time scales (scaling as a square root in the number
of particles). What we found in Sect.~\ref{sect:subtlety} is that some
quantum effects instead decouple by having a vanishing amplitude,
while their frequency scales as if they were classical effects.

Another interesting finding of our analysis is that in certain spin
systems the evolution does not lead to equilibration, even on the
incoherent time scales. For example, the system considered in
Sect.~\ref{sec:eigenstate_bg} is ``frozen-out'', unless the initial
numbers of spins up and down are very close. By studying this example
and comparing it to the case considered in
Sect.~\ref{sec:superposition_bg} we conclude the freeze-out happens in
the spin system for any mode where quantum effects are suppressed a
large classical (coherent) effect. We have seen, {\it e.g.}, in
Sect.~\ref{sec:freeze-out} that equilibration is precluded by the
conservation of the classical angular momentum. Only when the
classical angular momentum is reduced to the level comparable to the
quantum effects is the equilibration allowed to proceed. The special
case considered in  \cite{Friedland_&_Lunardini_1} falls into the
latter category.

We have explained in Sect.~\ref{sect:frezeout_musings} that the
physical origin of the freeze-out effect comes from dropping the
momentum degrees of freedom of the neutrino system. While the
freeze-out certainly happens in the spin system, it does not happen in
a real neutrino gas. This gives an important example of the
limitations of the spin model: while it captures all coherent
effects in the neutrino system, it does not always capture incoherent
effects, only those that are not classically suppressed.

In addition to giving the full solution in Eqs.~(\ref{eq:form of
  probability}), (\ref{eq:CJ}), and (\ref{eq:etaJ}), in several case
we have derived approximate expressions that elucidate the physics of
the system. For example, Eq.~(\ref{eq:coh_analyt}) cleanly
demonstrates the presence of both oscillation and decay in the spin
system.  Eq.~(\ref{eq:analyt_equil}) helps to understand the interplay
of classical and quantum effects in the processes of equilibration and
freeze-out. Finally,  Eq.~(\ref{eq:asympt_firstterm}) showed the
subtlety of decoupling the quantum effect in the classical limit.

We believe that our extensive investigation here will contribute to
better understanding of the flavor evolution in dense neutrino
systems. We also hope that our results could find applications beyond
the neutrino field.

\acknowledgments

A. F. was supported by the Department of Energy, under contract
W-7405-ENG-36. I.O.  and B.McK. were supported in part by the
Australian Research Council.

\appendix

\section{Explicit expressions for $C(J)$ and $\eta(J)$}
\label{sec:app_c_eta}

Eqs.~(\ref{eq:form of probability}), (\ref{eq:CJ}),
(\ref{eq:etaJ}) give the general form of the solution for the
probability $P_1(t)$. Here we give explicit expressions for $C(J)$
and $\eta(J)$ for the specific cases analyzed in the text.

In Sec.~\ref{sec:superposition_bg} we consider the initial state
containing $N$ spin up particles ($m_N=N/2$) and $M$ spins in the
orthogonal direction ($m_M=0$). In this case, Eqs.~(\ref{eq:CJ})
(\ref{eq:etaJ}), upon evaluating the Clebsch-Gordan, $3j$- and
$6j$-coefficients, become
\begin{eqnarray}
 \label{eq:C2_superposition}
 C(J) &=&
\frac{(1+2J)[\Delta(1+\Sigma)+J(1+J)]}
{4 J(1+J)\Gamma[1+J-N/2]\Gamma[1+J+\Delta]}\nonumber\\
&\times&\frac{\Gamma[1+J+\Delta]\Gamma[1+J+N/2]\Gamma[N+1]}
{\Gamma[\Sigma-J+1]\Gamma[\Sigma+J+2]},\\
 \label{eq:eta2_superposition}
\eta(J) &=& \frac{ \Gamma[2+J+\Delta]\Gamma[2+J+N/2]\Gamma[N]}
{(1+J)\Gamma[1+J-N/2]\Gamma[1+J+\Delta]}\nonumber\\
&\times&\frac{1}{\Gamma[\Sigma-J]\Gamma[\Sigma+J+2]}.
\end{eqnarray}
Here $\Delta\equiv\frac{N-M}{2}$ and $\Sigma\equiv\frac{N+M}{2}$.

Similarly, in Sec.~\ref{sec:eigenstate_bg} we consider the initial
state containing $N$ spin up particles ($m_N=N/2$) and $M$ spin
down particles ($m_M=-M/2$). In this case, Eqs.~(\ref{eq:CJ}),
(\ref{eq:etaJ}) evaluate to
\begin{eqnarray}
\label{eq:C2_eigenstates}
 C(J) &=&
\frac{(1+2J)M \Delta[\Delta(1+\Sigma)+J(1+J)]} {2 J(1+J)}\nonumber\\
&\times&
\frac{\Gamma(M)\Gamma(N)}{\Gamma[\Sigma-J+1]\Gamma[\Sigma+J+2]}
,\\
 \label{eq:eta2_eigenstates}
\eta(J) &=& \frac{M [(1+J)^2-\Delta^2)\Gamma(M)\Gamma(N)}
{(1+J)\Gamma[\Sigma-J]\Gamma[\Sigma+J+2]}.
\end{eqnarray}
Eq.~(\ref{eq:C2_eigenstates}) is ambiguous for $J=0$. In this
case, which occurs only when $N=M$, we have
\begin{eqnarray}
\label{eq:C2_eigenstates_zero}
 C(0) &=& 0.
\end{eqnarray}
Formally, this can be seen from Eq.~(\ref{eq:CJ}), in which both
the $3j$- and the $6j$-symbols contain objects that do not form
closed triangles and hence vanish.

Notice that our expressions agree with the solution given in F\&L
II \cite{Friedland_&_Lunardini_1} in the case $N=M$. Indeed,
setting $M=N$, $x=0$, and $y=N$ we have $C(J)=0$ and
$\eta(J)=(1+J) N [(N-1)!]^2/[(N-J-1)!(N+J+1)!]$. Noticed that
Eq.~(4.6) in the journal version of \cite{Friedland_&_Lunardini_1}
contains a typo: the factor $(1+2J)$ in the numerator should be
$(2+2J)$ (the first line of that equation is correct).

\section{Derivation of Eq.~(\ref{eq:coh_analyt})}
\label{sec:appendix_coh}

The derivation of Eq.~(\ref{eq:coh_analyt}) proceeds as follows.
Let us start with the definition of $\eta(J)$ in
Eq.~(\ref{eq:etaJ}). Approximating the gamma functions by the
Sterling formula, $\Gamma(z)=(z-1)!\simeq
\sqrt{2\pi}e^{(z-1/2)\ln(z-1)-(z-1)}$, and combining the resulting
exponents, we get
\begin{eqnarray}
\label{eq:appA_A}
\eta(J) &=& \frac{1}{(1+J)\sqrt{2\pi}} \exp[A],\nonumber\\
A &\equiv&
  \frac{1}{2} \left(
(3 + 2 J + M - N) \ln\left[\frac{2 + 2 J + M - N}{2}\right]
\right.\nonumber\\
&+& \left.(3 + 2 J + N) \ln\left[ \frac{2 + 2 J + N}{2}\right]
\right.\nonumber\\
&-& \left.( 2 J - N +1) \ln\left[ \frac{2J - N}{2}\right]  \right.\nonumber\\
&-& \left. (2 J - M + N +1) \ln\left[\frac{2 J - M + N}{2}\right]\right.\nonumber\\
&+& \left.(2 N-1) \ln[ N-1 ]
\right.\nonumber\\
&+& \left. (1 + 2 J - M - N) \ln\left[\frac{ -2 - 2 J + M +
N}{2}\right]
\right.\nonumber\\
&-& \left. (3 + 2 J + M + N) \ln\left[\frac{2 + 2 J + M +
N}{2}\right]\right.\nonumber\\
&-&\left.2\right).
\end{eqnarray}
We need to find the maximum of $A$ and expand $A$ in a series in
$J$ to the second order around this maximum. A derivative
$\partial A/\partial J$ has two types of terms: logarithms and
terms that go like $1/\#$, where ``\#'' denotes an expression
linear in $J$, $N$, and $M$. Clearly, for sufficiently large
values of $N$ and $M$ the logarithms dominate. The maximum is then
found by combining the logarithms and setting the argument of the
combined logarithm to 1. This yields
\begin{eqnarray}
  \label{eq:appA_log}
  \lefteqn{\frac{(2 + 2 J + M - N)(2 + 2 J + N)}{(2J - N)(2 J - M +
  N)}}\nonumber\\
  &\times&\frac{-2 - 2 J + M + N}{(2 + 2 J + M + N)}=1.
\end{eqnarray}
For large $N$ and $M$, we can drop small numbers compared to $J$,
$M$, $N$. The above equation then gives the answer
\begin{eqnarray}
  \label{eq:appA_min}
  J_{\rm max}=\frac{\sqrt{M^2+N^2}}{2},
\end{eqnarray}
which is exactly the answer for the classical problem.

Upon evaluating the second derivative, $\partial^2 A/\partial
J^2$, we again find two types of terms, some that go like $1/\#$
and some that go like $1/\#^2$. Keeping only the terms of the
first kind, substituting Eq.~(\ref{eq:appA_min}), and simplifying,
we get
\begin{eqnarray}
  \label{eq:appA_2nd_der}
  \partial^2 A/\partial J^2 = -\left(\frac{4}{N}+\frac{4N}{M^2} \right)
\end{eqnarray}

Finally, we need to find the value of $A$ in Eq.~(\ref{eq:appA_A})
for $J=J_{\rm max}$. This involves a rather lengthy calculation
involving a lot of cancellations.
All terms of the type $J \ln[\#]$, $N \ln[\#]$, and $M \ln[\#]$
cancel out, when Eq.~(\ref{eq:appA_log}) is used. The terms of the
type $\ln[\#]$ give $\ln(M^2/(4N))$ upon simplification. Finally,
the order one terms also cancel out. Thus, we get
\begin{equation}
  \label{eq:appA_AatJmax}
  A_{J_{\rm max}} = \ln\left(\frac{M^2}{4N}\right)
   + O\left(\frac{1}{N},\frac{1}{M},\frac{1}{J}\right),
\end{equation}
or
\begin{equation}
  \label{eq:appA_Aseries}
  A \simeq \frac{1}{2}\ln\left(\frac{M^2}{4N}\right)
  -\left(\frac{2}{N}+\frac{2N}{M^2} \right)(J-\sqrt{M^2+N^2}/2)^2
\end{equation}

This means that $\eta(J)$ is a Gaussian centered at $J_{\rm
  max}=\frac{\sqrt{M^2+N^2}}{2}$. For large $N$, $M$ the Gaussian is
sufficiently narrow and we can replace the prefactor $1/(1+J)$ by
$2/\sqrt{M^2+N^2}$. The sum over $J$ in Eq.~(\ref{eq:form of
probability}) can be replaced by an integral, in which we can
extend the limits of integration to $\pm\infty$. We have
\begin{widetext}
\begin{eqnarray}
  \label{eq:appA_integral}
  \lefteqn{\int_{-\infty}^{\infty} dJ \eta(J) \cos (2g J t) =}\nonumber\\
&&
\frac{2}{\sqrt{2\pi}\sqrt{M^2+N^2}}\left(\frac{M^2}{4N}\right)^{1/2}
\int_{-\infty}^{\infty} dJ e^{-\left(\frac{2}{N}+\frac{2N}{M^2}
\right)\left(J-\frac{\sqrt{M^2+N^2}}{2}\right)^2}
\cos (2g Jt) =\nonumber\\
&&
\frac{\sqrt{2}}{\sqrt{\pi}\sqrt{M^2+N^2}}\left(\frac{M^2}{4N}\right)^{1/2}
\frac{\sqrt{\pi}\sqrt{N M^2}}{\sqrt{2(M^2+N^2)}}
e^{-\left(\frac{2}{N}+\frac{2N}{M^2} \right)^{-1}g^2 t^2}
\cos (g\sqrt{M^2+N^2}t) =\nonumber\\
&&\frac{1}{2}\frac{ M^2}{(M^2+N^2)}
e^{-\left(\frac{2}{N}+\frac{2N}{M^2} \right)^{-1} g^2 t^2} \cos
(g\sqrt{M^2+N^2}t).
\end{eqnarray}
\end{widetext}

The constant term $1/2+ \sum_{J=J_{\rm min}}^{J_{\rm max}} C(J)$
can be immediately found as the difference between $1$ and the
oscillating term at $t=0$:
\begin{equation}
  1-\frac{1}{2}\frac{ M^2}{(M^2+N^2)}=
\frac{1}{2}+\frac{1}{2}\frac{ N^2}{(M^2+N^2)}.
\end{equation}
This concludes the derivation of  Eq.~(\ref{eq:coh_analyt}).

\section{Derivation of Eq.~(\ref{eq:analyt_equil})}
\label{sect:appB}

To find the constant value $\bar{P}_1$ to which $P_1(t)$ relaxes we
can either compute the sum $1/2+ \sum_{J=J_{\rm min}}^{J_{\rm max}}
\tilde{C}(J)$, or use the trick at the end of
Appendix~\ref{sec:appendix_coh} and compute $1-\sum_{J=J_{\rm
    min}}^{J_{\rm max}-1}\eta(J)$. Let us do the latter.

The first step is to approximate the gamma functions in
Eq.~(\ref{eq:eta2_eigenstates}) by the Sterling formula and expand
the exponent. After a fairly straightforward calculation, one finds
a Gaussian in $J+1$:
\begin{eqnarray}
  \label{eq:appB_gaussian}
  &&\frac{\Gamma(M+1)\Gamma(N)}
{\Gamma[(M+N)/2-J]\Gamma[(M+N)/2+J+2]} \simeq\nonumber\\
  && \exp \left[ B - \frac{2(M+N-1)}{(M+N)^2}(J+1)^2\right],
\end{eqnarray}
where
\begin{eqnarray}
  \label{eq:appB_gaussianB}
 B &\equiv& \left(M+\frac{1}{2}\right)\log M
+ \left(N-\frac{1}{2}\right)\ln(N-1) \nonumber\\
&-& (1+M+N)\ln\left(\frac{M+N}{2}\right) +1
\end{eqnarray}

For definitiveness, let us take $M>N$ and let us assume that $(M-N)
\ll N$. We can then expand Eq.~(\ref{eq:appB_gaussianB}) in series
in $\Delta$:
\begin{eqnarray}
  \label{eq:appB_seriesDelta}
  B &\simeq& \left(N-\frac{1}{2}\right)\ln(N-1) -
  \left(N+\frac{1}{2}\right)\log N\nonumber\\
  &+& 1 + \frac{2N-1}{8N^2}(N-M)^2
  \nonumber\\
  &=& -\ln N  + \frac{1}{4N}(N-M)^2 + O(1/N^2).
\end{eqnarray}

We have
\begin{eqnarray}
  \label{eq:appB_approxeta}
  \eta(J) &\simeq&
  \frac{[(1+J)^2-\Delta^2]}{(1+J)(\Delta+\Sigma)}
  \exp\left[\frac{\Delta^2}{\Delta+\Sigma} - \frac{(J+1)^2}{\Sigma}\right],\;\;\;\;
\end{eqnarray}
where, as before, $\Delta\equiv\frac{N-M}{2}$ and
$\Sigma\equiv\frac{N+M}{2}$.

Let us approximate the sum over $J$ by an integral
\begin{eqnarray}
  \label{eq:appB_sumint}
  \lefteqn{\bar{P}_1=1-\sum_{J=(M-N)/2}^{(M+N)/2-1}\eta(J)
   }\nonumber\\
&\rightarrow&  1-\int_{\Delta}^{\Sigma-1} dJ
  \frac{(J^2-\Delta^2)}{J (\Delta+\Sigma)}
  \exp\left[\frac{\Delta^2}{\Delta+\Sigma} - \frac{J^2}{\Sigma}\right].
\end{eqnarray}
Let us shift the integration variable $\tilde{J}= J -
\Delta$ and extend the upper limit of integration to infinity:
\begin{eqnarray}
  \label{eq:appB_shift}
  &&1-\bar{P}_1\simeq
\\
  &&\int_{0}^{\infty} d\tilde{J}
  \frac{(\tilde{J}+2\Delta)\tilde{J}}{(\tilde{J}+\Delta) (\Delta+\Sigma)}
  \exp\left[\frac{\Delta^2}{(\Delta+\Sigma)}
    - \frac{ (\tilde{J}+\Delta)^2}{\Sigma}\right].\nonumber
\end{eqnarray}

Introducing a new variable $\xi=\tilde{J}/\sqrt{\Sigma}$, we get
\begin{eqnarray}
  \label{eq:appB_rescale}
  1-\bar{P}_1\simeq\frac{\Sigma}{\Delta+\Sigma}
\exp\left[\frac{-\Delta^3}{(\Delta+\Sigma) \Sigma}\right]
  F\left(\frac{\Delta}{\sqrt{\Sigma}} \right),
\end{eqnarray}
where
\begin{eqnarray}
  \label{eq:appB_defF}
  F(a)\equiv   \int_{0}^{\infty} d\xi \frac{(\xi+2 a)\xi}{(\xi+a)}
  \exp\left[  -  (\xi^2+2\xi a)\right]
\end{eqnarray}

The function $F(a)$ drops to zero quickly as $a$ is increased beyond
one. Since the argument of $F$ in Eq.~(\ref{eq:appB_rescale}) is
$\Delta/\sqrt{\Sigma}$, we conclude that indeed
$\bar{P}_1\rightarrow 1$ for $\Delta\gtrsim \sqrt{\Sigma}$.

For $\Delta\lesssim \sqrt{\Sigma}$ the argument of the exponent in
Eq.~(\ref{eq:appB_rescale}) is of the order $\Delta^3/\Sigma^2
\lesssim \Sigma^{-1/2}$ and hence in the limit of a large number of
spins the exponent can be set to one.

The integral defining the function $F(a)$ can be further transformed
as follows:
\begin{eqnarray}
  \label{eq:appB_trasform_F}
    F(a) &=& \int_{0}^{\infty} d\xi \frac{(\xi+a)^2-a^2}{\xi+a}
    \exp\left[  -  (\xi+ a)^2\right] e^{a^2}\nonumber\\
&=&e^{a^2}\int_{0}^{\infty} d\xi
\left((\xi+a)-\frac{a^2}{\xi+a}\right)
    \exp\left[  -  (\xi+ a)^2\right] \nonumber\\
&=& \frac{1}{2}-\frac{e^{a^2}}{2} \int_{0}^{\infty} d(\xi+a)^2
    \frac{a^2}{(\xi+a)^2}
    \exp\left[  -  (\xi+ a)^2\right]\nonumber\\
&=& \frac{1}{2}-\frac{a^2}{2} \exp[a^2]\int_{a^2}^{\infty} dy
    \frac{e^{-  y}}{y}
\end{eqnarray}

Finally, we arrive at
\begin{eqnarray}
  \label{eq:appB_final}
  \bar{P}_1 \simeq 1 - \frac{\Sigma}{\Delta+\Sigma}
  F\left(\frac{\Delta}{\sqrt{\Sigma}} \right),
\end{eqnarray}
where
\begin{eqnarray}
  \label{eq:appB_final1}
   F(a) = \frac{1}{2}-\frac{a^2}{2} \int_{0}^{\infty} dy
    \frac{e^{-  y}}{y+a^2}.
\end{eqnarray}

The function $F(a)$ equals $1/2$ at $a=0$
($F(a)\stackrel{a\rightarrow 0}{=}1/2+1/2(2\ln a+\gamma)a^2 +1/2
(2\ln a+\gamma-1)a^4 + ... $, $\gamma\simeq0.577216$ being the Euler
constant). As $a\rightarrow\infty$ $F(a)$ approaches $0$
($F(a)\stackrel{a\rightarrow\infty}{=}a^{-2}/2+...$).

\section{Derivation of Eq.~(\ref{eq:asympt_firstterm})}
\label{sect:app_asympt_series}

In Sect.~\ref{sec:eigenstate_bg} we consider the evolution of the
spin system in which all spins are initially in flavor eigenstates.
In this Appendix we derive an approximate expression describing this
evolution in the regime when $\sqrt{M+N}<|M-N|\ll(M+N)$.

The time dependent part of the evolution is given by $\sum_J \eta
(J)\cos(2gt(J+1))$. The coefficient $\eta(J)$ depends on
$\Delta=(N-M)/2$ and can be usefully approximated for the case when
$\Delta$ is small by Eq.~(\ref{eq:appB_approxeta}) of Appendix
\ref{sect:appB}.

As before, we approximate the sum by an integral
\begin{eqnarray}
  \label{eq:app_asymp_series_sumint}
  \lefteqn{\sum_{J=(M-N)/2}^{(M+N)/2-1}\eta(J)\cos(2gt(J+1))
   }\\
&\rightarrow&  \int_{\Delta}^{\Sigma-1} dJ
  \frac{(J^2-\Delta^2)}{J (\Delta+\Sigma)}
  \exp\left[\frac{\Delta^2}{\Delta+\Sigma} - \frac{J^2}{\Sigma}\right]\cos(2gt J),\nonumber
\end{eqnarray}
shift the variable, $J$ to $\tilde{J}=J-\Delta$, and extend the
range of integration to infinity. We find
\begin{widetext}
\begin{eqnarray}
    \label{eq:app_asymp_series_starting_int}
  \sum_{J=(M-N)/2}^{(M+N)/2-1}\eta(J)\cos(2gt(J+1))
&\simeq& \exp\left[\frac{-\Delta^3}{(\Delta+\Sigma)\Sigma}\right]\frac{1}{(\Delta+\Sigma)}\nonumber\\
&\times&\int_{0}^{\infty} dJ
  \frac{(\tilde{J}+2\Delta)\tilde{J}}{(\tilde{J}+\Delta) }
  \exp\left[- \frac{\tilde{J}^2}{\Sigma}-\frac{2\tilde{J}\Delta}{\Sigma}\right]
  \cos(2gt (\tilde{J}+\Delta)).
\end{eqnarray}
\end{widetext}
F\&L II considered a limit $\Delta=0$. In that case the integral
equals
\begin{eqnarray}
    \label{eq:app_asymp_series_FL_int}
    \frac{1}{2}-\frac{\sqrt{\pi}}{2}\sqrt{N}g t\exp(-N g^2 t^2)
    {\rm erfi}(\sqrt{N}g t),
\end{eqnarray}
from which Eq.~(\ref{eq:FLIIintegral}) immediately follows.

We notice that the values of $\tilde{J}$ that contribute in the case
$\Delta=0$ are of the order $\sqrt{\Sigma}$, as the integral in
Eq.~(\ref{eq:app_asymp_series_starting_int}) is cut off by the term
$\exp(-\tilde{J}^2/\Sigma)$.  If we take $\Delta>\sqrt{\Sigma}$,
however, we find that the integral is first cut off by the exponential
term $\exp(-2\tilde{J}\Delta/\Sigma)$, at $\tilde{J}\sim
\Sigma/2\Delta$. If $\Delta$ is sufficiently large, we may evaluate
the integral in Eq.~(\ref{eq:app_asymp_series_starting_int}) with the
first term in the exponent dropped. The integral then becomes
\begin{eqnarray}
    \label{eq:app_asymp_series_int_droppedJ2}
 \int_{0}^{\infty} d\tilde{J}
  \frac{(\tilde{J}+2\Delta)\tilde{J}}{(\tilde{J}+\Delta) }
  \exp\left[-\frac{2\tilde{J}\Delta}{\Sigma}\right]
  \cos(2gt (\tilde{J}+\Delta)).
\end{eqnarray}
We next take advantage of the fact that integrals of the form
$\int_0^\infty dx\;x^n\exp[-x]\cos(x+\alpha)$ can be done and the
answer has a simple and useful analytic form
\begin{eqnarray}
    \label{eq:app_asymp_series_int_ansatz}
 &&\int_{0}^{\infty} dx x^n \exp\left[-A x\right] \cos[B
 (x+\alpha)]=
 \nonumber\\
 &&\frac{n!}{(A^2+B^2)^{\frac{n+1}{2}}}\cos\left[B\alpha+(1+n)\arctan\left(\frac{B}{A}\right)\right].
  \;\;\;\;\;
\end{eqnarray}
Expanding $\frac{\tilde{J}+2\Delta}{\tilde{J}+\Delta}=
2-\frac{\tilde{J}}{\Delta}+\frac{\tilde{J}^2}{\Delta^2}-\frac{\tilde{J}^3}{\Delta^3}+...$,
we can reduce the integral in
Eq.~(\ref{eq:app_asymp_series_int_droppedJ2}) to a series of
integrals of the form (\ref{eq:app_asymp_series_int_ansatz}).
Evaluating this series, we obtain
\begin{widetext}
\begin{eqnarray}
    \label{eq:app_asymp_series_int_answer}
&&I(t)\equiv\int_{0}^{\infty} d\tilde{J}
  \frac{(\tilde{J}+2\Delta)\tilde{J}}{(\tilde{J}+\Delta) }
  \exp\left[-\frac{2\tilde{J}\Delta}{\Sigma}\right]
  \cos(2gt (\tilde{J}+\Delta))\simeq\nonumber\\
  &&\frac{1}{2}
  \left(g^2t^2+\frac{\Delta^2}{\Sigma^2}
  \right)^{-1}
  \cos\left[2 g t \Delta+2\arctan\left(\frac{g t
  \Sigma}{\Delta}\right)\right]+
  \nonumber\\
  &&\sum_{n=2}^{n_{\rm max}} (-1)^{n+1}\frac{n!}{2^{n+1}} \Delta^{1-n} \left(g^2t^2+\frac{\Delta^2}{\Sigma^2}
  \right)^{-(n+1)/2}
  \cos\left[2 g t \Delta+(1+n)\arctan\left(\frac{g t
  \Sigma}{\Delta}\right)\right].
  \end{eqnarray}
\end{widetext}

The form of Eq.~(\ref{eq:app_asymp_series_int_answer}) suggests that
the solution is oscillating in time. It is instructive to understand
how this oscillatory behavior disappears in the limit
$\Delta^2/\Sigma\rightarrow 0$. First of all, notice that the
summation in Eq.~(\ref{eq:app_asymp_series_int_answer}) goes to some
$n_{\rm max}$, \emph{not to infinity}. Indeed, the presence of the
factorial $n!$ indicates that we are dealing with an asymptotic
series. For any given choice of $\Delta$ and $\Sigma$ there will be an
optimal number of terms in the series, $n_{\rm max}$, that
approximates the original integral best. The terms beyond $n_{\rm
  max}$ grow in absolute value and the series diverges. We easily
estimate $n_{\rm max}$ from the condition that the ratio of the two
consecutive terms at $t=0$ be $\sim 1$,
$(n/2)\Delta^{-1}\Delta^{-1}\Sigma\sim 1$, or
\begin{equation}\label{eq:n_max}
    n_{\rm max} \sim 2\Delta^2/\Sigma.
\end{equation}
Clearly, for $\Delta^2\lesssim\Sigma$ the expansion breaks down.

Next, recall that in this limit the integral in
Eq.~(\ref{eq:app_asymp_series_int_droppedJ2}) is not valid anyway
(the first term in the exponent in
Eq.~(\ref{eq:app_asymp_series_starting_int}) cannot be dropped) and
one needs to consider Eq.~(\ref{eq:app_asymp_series_starting_int}).
The answer to the latter in the limit $\Delta^2/\Sigma\rightarrow 0$
is provided by Eq.~(\ref{eq:app_asymp_series_FL_int}), indeed not
showing any oscillations. The oscillations thus appear as $|M-N|$ is
increased beyond $\sqrt{M+N}$, when the series in
Eq.~(\ref{eq:app_asymp_series_int_answer}) starts providing a better
and better description of the true answer.

The final answer for the probability $P_1(t)$ in the regime
$\sqrt{M+N}<|M-N|\ll(M+N)$ is given by
\begin{eqnarray}
    \label{eq:app_asymp_series_answer}
P_1(t)&=&1-\sum_J \eta (J) + \sum_J \eta (J)\cos(2gt(J+1))\nonumber\\
  &\simeq&1+\frac{\exp\left[\frac{-\Delta^3}{(\Delta+\Sigma)\Sigma}\right]}{(\Delta+\Sigma)}(I(t)-I(0)),
  \end{eqnarray}
where $I(t)$ is given by the series in
Eq.~\ref{eq:app_asymp_series_int_answer}. Taking the leading term in
the series and setting $\Delta=0$ in the prefactor of
Eq.~(\ref{eq:app_asymp_series_answer}) we arrive at
Eq.~(\ref{eq:asympt_firstterm}). This approximation turns out to be
quite accurate, as Fig.~\ref{fig:asymptotic}, which shows the case
$N=2300$, $M=2700$, illustrates.

\end{document}